\documentclass[iop, apj]{emulateapj}

\newcommand{\kms}{\rm km~s^{-1}}
\newcommand{\kmsmpc}{\rm km~s^{-1}~Mpc^{-1}}

\slugcomment{\bf Last updated:\today}

\usepackage{color}
\usepackage{multirow}

\begin{document}

\title{The HectoMAP Cluster Survey - I. redMaPPer Clusters}

\author{Jubee Sohn$^{1}$,
        Margaret J. Geller$^{1}$, 
        Kenneth J. Rines$^{2}$,
        Ho Seong Hwang$^{3}$,
        Yousuke Utsumi$^{4}$, 
        Antonaldo Diaferio$^{5,6}$} 

\affil{$^{1}$ Smithsonian Astrophysical Observatory, 60 Garden Street, Cambridge, MA 02138, USA}
\affil{$^{2}$ Department of Physics \& Astronomy, Western Washington University, Bellingham, WA 98225, USA}
\affil{$^{3}$ Quantum Universe Center, Korea Institute for Advanced Study, 85 Hoegiro, Dongdaemun-gu, Seoul 02455, Korea}
\affil{$^{4}$ Kavli Institute for Particle Astrophysics and Cosmology, SLAC National Accelerator Laboratory, Stanford University, SLAC, 2575 Sand Hill Road, M/S 29, Menlo Park, CA  94025, USA} 
\affil{$^{5}$ Universit{\`a} di Torino, Dipartimento di Fisica, Torino, Italy}
\affil{$^{6}$ Istituto Nazionale di Fisica Nucleare (INFN), Sezione di Torino, Torino, Italy}

\begin{abstract}
We use the dense HectoMAP redshift survey 
 to explore the properties of 104 redMaPPer cluster candidates. 
The redMaPPer systems in HectoMAP  
 cover the full range of richness and redshift ($0.08 < z < 0.60$). 
Fifteen systems included in the {\it Subaru}/Hyper Suprime-Cam public data release
 are {\it bona fide} clusters. 
The median number of spectroscopic members per cluster is $\sim20$. 
We include redshifts of 3547 member candidates 
 listed in the redMaPPer catalog whether they are cluster members or not. 
We evaluate the redMaPPer membership probability spectroscopically.
The scaled richness ($\lambda_{rich}/S$) provided by redMaPPer 
 correlates tightly with the spectroscopically corrected richness 
 regardless of the cluster redshift.
The purity (number of real systems) in redMaPPer exceeds 90\% even at the lowest richness; 
 however, there is some incompleteness. 
Three massive galaxy clusters ($M \sim 2 \times 10^{13} M_{\odot}$) associated with X-ray emission 
 in the HectoMAP region are missing from the public redMaPPer catalog with $\lambda_{rich} > 20$.
\end{abstract}
\keywords{cosmology: observations -- large-scale structure of universe -- galaxies: clusters: general -- galaxies: clusters}
\section{INTRODUCTION}

Galaxy clusters are important probes
 of the formation and evolution of large scale structure in the universe. 
The cluster luminosity function  and the masses of galaxy clusters 
 provide strong constraints on model for the development of  large scale structure.
Beginning with Abell \citep{Abell58, Abell89} and Zwicky \citep{Zwicky68} 
 large cluster candidate catalogs have been 
 based on various techniques including optical catalogs 
 (e.g. \citealp{Gladders00, Koester07, Rykoff14, Oguri17}), 
 X-ray samples (e.g. \citealp{Edge90, Gioia90, Ebeling98, Bohringer00, Bohringer01, Bohringer04, Ebeling10, Pacaud16, Bohringer17}), 
 Sunyaev-Zel'dovich samples (e.g. \citealp{Melin06, Vanderlinde10, Marriage11, Bleem15, Planck15, Planck16}),
 and weak lensing samples\citep{Oguri17}.

Many cluster surveys use sophisticated techniques along with photometric redshifts 
 to construct robust cluster candidate catalogs and
 to avoid systematic biases
 \citep{Koester07, Wen09, Wen12, Hao10, Szabo11, Oguri14, Rykoff14, Durret15, Rykoff16, Oguri17}.
These surveys identify cluster candidates 
 based on characteristic features of clusters including
 overdensities on the sky, identification of the brightest cluster galaxy, and sampling of the red-sequence defined by potential cluster members.
Generally these catalogs determine 
 cluster membership based on photometric redshifts of individual galaxies. 
The photometric redshifts remove some but not all chance alignments. 
The typical error in a photometric redshift is generally large 
 compared with the typical velocity dispersion of even the most massive clusters. 
 
Here we compare a photometrically selected sample, redMaPPer \citep{Rykoff14, Rykoff16}, 
 with the dense redshift survey, HectoMAP.
redMaPPer (The red-sequence Matched-filter Probabilistic Percolation) 
 is a red-sequence cluster finding survey 
 covering the Sloan Digital Sky Survey (SDSS) Data Release 8 (DR8) data. 
redMaPPer identifies the red-sequence of galaxies
 with the guidance of photometric redshifts.
The redMaPPer catalog provides an important testbed for these identification algorithm
 because it includes membership probabilities of individual galaxies along with a cluster richness. 
The richness is a potential mass proxy 
 that has been tested with shallower redshift surveys \citep{Rozo15b, Rines17}. 
Here we test the full redshift and richness range of the catalog. 

Previous tests of redMaPPer include 
\citet{Rozo14} and \citet{Sadibekova14} 
 who examine the properties of redMaPPer clusters 
 coincident with X-ray and SZ cluster candidates. 
\citet{Rozo15a} compare the redMaPPer and 
 the {\it Planck} SZ cluster candidate catalog. 
\citet{Rozo15b}  
 compare the redMaPPer photometric membership probability estimate with spectroscopically
 determined membership from the SDSS and Galaxy and Mass Assembly (GAMA) surveys. 
They suggest that 
 there is a small ($\sim 2.4\%$) systematic offset 
 between the redMaPPer membership probability and the spectroscopic assessment. 
They also find only a small contamination of 
 the richness by non-cluster galaxies. 
These comparisons are largely restricted to $z \lesssim 0.3$.

\citet{Rines17} examine the spectroscopic properties of 23 high-richness redMaPPer clusters 
 in the redshift range 0.08 to 0.25 based on dense cluster redshift surveys 
 including $\sim 75$ members per cluster. 
In contrast with \citet{Rozo15b},
 their spectroscopic membership identification shows 
 that the redMaPPer membership probability is substantially overestimated for high-probability members
 and is substantially underestimated for low-probability members. 
In spite of this disagreement, 
 the redMaPPer richness is well-correlated with the velocity dispersion derived from the spectroscopy. 

HectoMAP \citep{Geller05, Geller11, Hwang16} is a unique sample
 for examining the spectroscopic properties of redMaPPer clusters throughout the redshift range they cover.
Here, we study 
 the 104 redMaPPer cluster candidates in the 53 deg$^{2}$ HectoMAP field. 
We examine the purity and  completeness of redMaPPer catalog 
 based on the dense HectoMAP spectroscopy.

We describe the HectoMAP redshift survey 
 and the redMaPPer cluster sample in Section 2. 
We investigate the spectroscopic properties of the redMaPPer clusters including the accuracy of the photometric cluster redshifts 
in Section 3. 
We discuss the implications 
 of the spectroscopic study of photometrically identified clusters
 including the photometric richness and completeness of the redMaPPer catalog in Section 5. 
We summarize in Section 6. 
We use the standard $\Lambda$CDM cosmology with 
 H$_{0} = 70~\kmsmpc$, $\Omega_{m} = 0.3$, $\Omega_{\Lambda} = 0.7$
 throughout this paper.

\section{DATA}

In Section \ref{hectomap}, we describe the dense redshift survey HectoMAP. 
We review the photometrically identified clusters in the HectoMAP region in Section \ref{redmapper}. 

\subsection{HectoMAP}\label{hectomap}

HectoMAP is a dense redshift survey with a median redshift, $z = 0.39$. 
The average number density of galaxies with spectroscopic redshifts is  
 $\sim2000$ deg$^{-2}$ ($\sim1200$ galaxies deg$^{-2}$ are in the highly complete red-selected subsample, 
 \citealp{Geller11, Geller15, Hwang16, Sohn17c}).
HectoMAP covers 52.97 deg$^{2}$ within the boundaries
 $200 < $ R.A. (deg) $ < 250$ and  $42.5 < $ Decl. (deg) $< 44.0$. 
The Sloan Digital Sky Survey (SDSS) Data Release 9 (DR9) photometric catalog \citep{Ahn12}
 is the photometric basis for the survey. 
The primary survey targets are red galaxies with 
 $(g-r)_{model, 0} > 1.0$, $(r-i)_{model, 0} > 0.5$, $r_{petro, 0} < 21.3$, and $r_{fiber, 0} < 22.0$ galaxies.
The color selection efficiently filters out galaxies with $z \lesssim 0.2$
 where the SDSS Main Galaxy Sample is reasonably dense. 
The $r_{fiber, 0}$ selection removes low surface brightness galaxies that are beyond the limit of our spectroscopy. 

We measured redshifts with the 300-fiber spectrograph
 Hectospec mounted on MMT 6.5m telescope \citep{Fabricant98, Fabricant05}
 from 2009 to 2016. 
The Hectospec yields $\sim250$ spectra within a $\sim1$ deg${^2}$ field of view 
 in a single observation. 
We used the 270 mm$^{-1}$ grating yielding a wavelength coverage of 3700 -- 9150 \AA~
 with a resolution of 6.2 \AA. 
The typical exposure time for an observation is 0.75 - 1.5 hr; 
 each observation is composed of three subexposures for cosmic ray removal. 
We used the HSRED v2.0 package
 originally developed by R.Cool and revised by the SAO Telescope Data Center (TDC) staff to reduce the data. 
We measured the redshifts by applying the cross-correlating code RVSAO \citep{Kurtz98}. 
Based on visual inspection of each spectrum, 
 we classified redshifts into three categories:
 high quality spectra (Q), ambiguous fits (?), and poor fits (X). 
We use only redshifts with `Q' for scientific analyses.
We obtained 58211 redshifts for red galaxies with $r_{petro, 0} < 21.3$ and 
 a total of 97929 redshifts (with no color selection) in the HectoMAP region. 
The internal redshift error normalized by $(1 + z)$ is $\sim 32~\kms$. 

\begin{figure}
\centering
\includegraphics[scale=0.49]{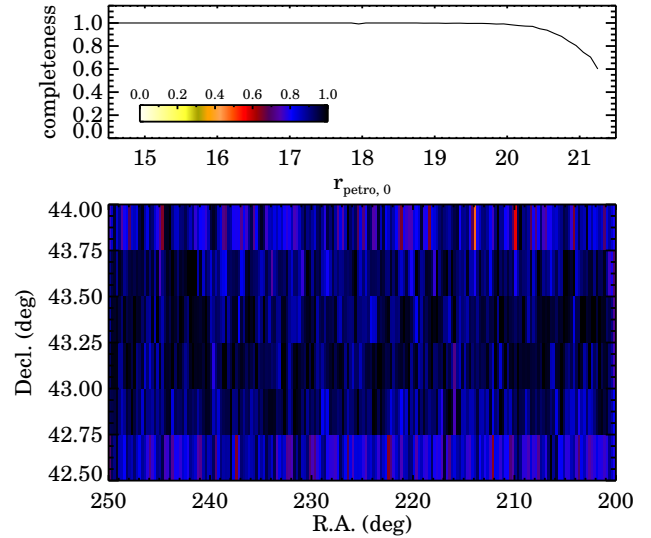}
\caption{(Upper) Spectroscopic survey completeness for HectoMAP galaxies
 as a function of $r-$band magnitude.
(Lower) Two-dimensional completeness map ($200 \times 6$ pixels) of HectoMAP
 for galaxies with $r_{petro, 0} \leq 21.3$,
 $(g-r)_{fiber, 0} > 1.0$, and $(r-i)_{fiber, 0} > 0.5$.}
\label{survey}
\end{figure}

HectoMAP is remarkably complete within the red galaxy selection limits:
 the survey is 98\% complete to $r_{petro, 0} < 20.5$ and 
 it is 89\% complete to $r_{petro, 0} < 21.3$.
Figure \ref{survey} shows the two-dimensional completeness map for 
 HectoMAP red galaxies. 
The coverage is uniform over the entire survey region. 
These objects are the main galaxies that enter in the evaluation of the redMaPPer algorithm for cluster identification. 

Outside the red color selection, the survey completeness is patchy. 
We use the bluer galaxies to maximize the number of galaxies that are potential redMaPPer cluster members. 
Below $z \sim 0.2$ the SDSS Main Sample is the primary redshift source of potential redMaPPer cluster members. 
The SDSS is also uniform over the HectoMAP region and 
 the average completeness regardless of color is $\gtrsim 90\%$.
 
Public {\it Subaru}/Hyper Suprime Cam (HSC) imaging covers $\sim 7$ deg$^{2}$ of the HectoMAP region. 
Eventually the entire region will be covered \citep{Aihara17}. 
We use the public images in Section \ref{result} as a partial test of the redMaPPer algorithm. 
In Sections \ref{spec}, 
 we highlight the properties of the 15 redMaPPer cluster candidates imaged with {\it Subaru}. 

\subsection{Photometrically Identified Cluster Catalogs: redMaPPer}\label{redmapper}

Many studies identify galaxy cluster candidates 
 based on photometric measures  
 including the red-sequence, 
 the over-density based on photometric redshifts, 
 and the identification of red objects associated with weak lensing peaks
 \citep{Koester07, Wen09, Wen12, Hao10, Szabo11, Oguri14, Rykoff14, Rykoff16, Durret15, Oguri17}.
The SDSS plays an important role in these photometric cluster surveys 
 thanks to its wide sky coverage. 
Several previous catalogs based on the SDSS include cluster candidates within the HectoMAP field.

We first summarize the numbers of 
 previous photometrically identified clusters
 within the HectoMAP region. 
Table \ref{hmap_stat} lists the number of cluster candidates in HectoMAP
 from MaxBCG \citep{Koester07}, GMBCG \citep{Hao10}, AMF \citep{Szabo11}, 
 WHL \citep{Wen09, Wen12}, CAMIRA \citep{Oguri14}, and redMaPPer \citep{Rykoff14, Rykoff16}.
For WHL and redMaPPer (v6.3), 
 we use the most recent versions from \citet{Wen12} and \citet{Rykoff16}, 
 respectively. 
 
\begin{deluxetable}{lccc}
\tablecolumns{4}
\tabletypesize{\scriptsize}
\tablewidth{0pt}
\tablecaption{Photometrically Identified Cluster Candidates in HectoMAP}
\tablehead{
\colhead{Catalogs} & \colhead{$N_{cand}$\tablenotemark{*}} & \colhead{$z$ range} & \colhead{ref.}}
\startdata
MaXBCG    & 133 & $0.12 < z < 0.30$ & \citet{Koester07} \\
GMBCG     & 361 & $0.13 < z < 0.54$ & \citet{Hao10}     \\
AMF       & 421 & $0.06 < z < 0.67$ & \citet{Szabo11}   \\
WHL       & 544 & $0.08 < z < 0.74$ & \citet{Wen12}     \\ 
CAMIRA    & 285 & $0.11 < z < 0.60$ & \citet{Oguri14}   \\
redMaPPer & 104 & $0.09 < z < 0.60$ & \citet{Rykoff16}
\enddata
\tablenotetext{*}{Number of cluster candidates within the HectoMAP area. }
\label{hmap_stat}
\end{deluxetable}  
 
The number of cluster candidates in the HectoMAP region 
 varies from 104 to 544.
The number of cluster candidates depends in part on 
 the cluster identification algorithm, 
 the limiting survey redshift, and the richness range of candidate clusters. 
Table \ref{hmap_stat} shows that 
 cluster surveys covering wider redshift ranges
 tend to identify more cluster candidates as expected. 
Within a fixed redshift range, 
 a cluster catalog including low richness candidates 
 contains a larger number of candidate systems.
However, we do not compare the richness distributions of the various catalogs,
 because the definitions of richness vary substantially. 

As a test of photometric catalogs, we focus on redMaPPer, 
 a catalog that has already been compared with X-ray observations \citep{Rozo14}, 
 SZ observations \citep{Rozo14, Rozo15a}, and weak lensing  \citep{Simet17}. 
The catalog has also been compared with 
 the SDSS and GAlaxy and Mass Assembly (GAMA, \citealp{Driver09}) spectroscopic surveys \citep{Rozo15b}. 
In contrast with HectoMAP, GAMA has a median redshift of $\sim0.2$ \citep{Hopkins13}. 
\citet{Rines17} have made a detailed test of the redMaPPer algorithm for 
 a set on nearby rich clusters with redshifts $0.08 < z < 0.25$. 

HectoMAP allows extension of the tests of redMaPPer to the catalog limit, $z \sim 0.6$. 
Conveniently, the magnitude limit of HectoMAP ($r_{petro, 0} = 21.3$) at $z = 0.5$ corresponds to
 $M_{r} = -20.96$ comparable with the $L_{*}$ of massive clusters \citep{Sohn17a}. 
Thus HectoMAP contains only redshifts of the brightest few cluster members 
 for candidates at $z > 0.5$. 
However, even at $z > 0.5$ HectoMAP provides a test of the
 redMaPPer membership probability assignments. 
 
\begin{figure}
\centering
\includegraphics[scale=0.49]{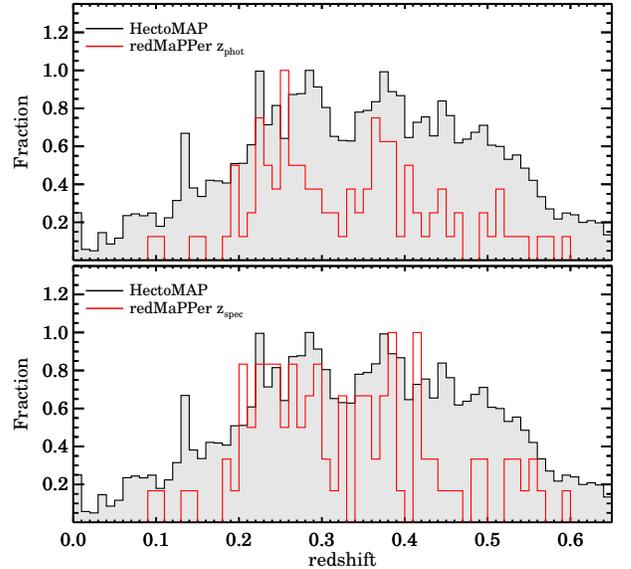}
\caption{
(Upper) Spectroscopic redshift distribution of individual HectoMAP targets (black filled histogram) and 
 the photometric redshift distribution of HectoMAP-red clusters from redMaPPer (red open histogram, 104 systems).
(Lower) Same as the upper panel, but the spectroscopic redshift distribution 
 of HectoMAP-red clusters from this paper (red open histogram).}
\label{zhist}
\end{figure}
 
As a first view of the relationship between the redMaPPer cluster sample and HectoMAP, 
 the upper panel of Figure \ref{zhist} compares 
 the redshift distribution of HectoMAP galaxies 
 with the distribution of redMaPPer cluster photometric redshifts. 
The peaks of the two redshift distributions are not coincident. 
For example, HectoMAP has its maximum at $z \sim 0.28$, 
 but the corresponding redMaPPer peak is at lower redshift. 
HectoMAP red galaxies are abundant at higher redshift, $z > 0.4$, 
 where redMaPPer identifies few candidates. 
The lower panel of Figure \ref{zhist} shows that 
 the difference diminishes when we plot the distribution of spectroscopic redshifts 
 of redMaPPer clusters (determined in Section 3).

\section{redMaPPer Clusters in HectoMAP}\label{result}

There are 104 redMaPPer cluster candidates in the HectoMAP region. 
Hereafter, we refer to these cluster candidates as HectoMAP-red clusters. 
The redshift of HectoMAP-red cluster sample is $0.08 < z < 0.60$, and 
 the richness (redMaPPer $\lambda_{rich}$ parameter) ranges from 20 to 106.
The redshift and richness ranges for HectoMAP-red clusters 
 are distinctive in covering the full ranges of the redMaPPer cluster candidate catalog. 

We summarize the HectoMAP-red clusters in Table \ref{hmapred}. 
We list the redMaPPer ID, R.A., Decl., redMaPPer richness ($\lambda_{rich}$),  
 photometric redshift ($z_{phot}$), HectoMAP spectroscopic redshift ($z_{spec}$), 
 the number of redMaPPer members with $P_{mem} > 0$ (N$_{RM, mem}$), 
 the number of spectroscopically identified members (N$_{Spec, mem}$), 
 the spectroscopic completeness ($f_{RM, mem}$), 
 the spectroscopically identified member fraction among redMaPPer members ($f_{Spec-mem, cl}$, Section \ref{cont}). 
Table \ref{hmapredmem} also lists 
 the SDSS object ID, R.A., Decl., redMaPPer membership probability, redshift, source of redshift, and spectroscopic membership
 for the 3547 redMaPPer objects with $P_{mem} > 0$ with HectoMAP redshift. 
Here we refer to the total membership probability as the redMaPPer membership probability
 (see the Note for Table 7 in \citealp{Rykoff16}).  
The total membership probability is the radius and luminosity weighted probability: 
 $P_{mem} = P \times P_{free} \times \theta_{I} \times \theta_{R}$, 
 where $P$ is raw membership probability, 
 $P_{free}$ is the probability that member is not a member of a higher ranked cluster, 
 $\theta_{i}$ is the $i-$band luminosity weight, 
 and $\theta_{r}$ is the radial weight. 
Table \ref{hmapredmem} includes objects whether or not they are spectroscopically identified members.  

\begin{deluxetable*}{lccccccccc}
\tablecolumns{10}
\tabletypesize{\scriptsize}
\tablewidth{0pt}
\tablecaption{HectoMAP-red clusters}
\tablehead{
\colhead{Cluster ID} & \colhead{R.A.} & \colhead{Decl.} & 
\colhead{$\lambda_{rich}$\tablenotemark{a}} & 
\colhead{$z_{phot}$\tablenotemark{a}} &  
\colhead{$z_{spec}$\tablenotemark{b}} & 
\colhead{N$_{RM, mem}$\tablenotemark{a}} & 
\colhead{N$_{Spec-mem}$\tablenotemark{b}} & 
\colhead{$f_{comp}$\tablenotemark{c}} & 
\colhead{$f_{spec-mem, cl}$\tablenotemark{d}} }
\startdata
05570 & 14:13:43.5 & 43:38:41 &  25.37 & 0.0901 & 0.0894 &  31 &  51 & 0.94 & 0.90 \\
05706 & 14:17:54.2 & 43:23:16 &  27.66 & 0.1054 & 0.1060 &  42 &  41 & 0.81 & 0.76 \\
08065 & 16:21:26.9 & 42:45:40 &  26.55 & 0.1424 & 0.1380 &  40 &  43 & 0.43 & 0.88 \\
09448 & 15:32:39.7 & 43:03:28 &  22.29 & 0.1542 & 0.1430 &  35 &  24 & 0.49 & 0.82 \\
03312 & 16:26:42.5 & 42:40:11 &  41.28 & 0.1871 & 0.1870 &  56 &  54 & 0.64 & 0.89 \\
09234 & 16:26:23.8 & 42:53:20 &  24.92 & 0.1913 & 0.1867 &  52 &  42 & 0.81 & 0.50 \\
13503 & 16:19:18.4 & 42:46:10 &  23.37 & 0.1959 & 0.1936 &  38 &   8 & 0.34 & 0.31 \\
09874 & 15:12:57.4 & 43:18:41 &  22.25 & 0.1978 & 0.2059 &  50 &  26 & 0.54 & 0.52 \\
18313 & 15:12:18.7 & 43:33:14 &  21.86 & 0.1983 & 0.2095 &  39 &  20 & 0.49 & 0.53 \\
07253 & 14:51:29.5 & 42:35:34 &  33.58 & 0.2098 & 0.2044 &  52 &  13 & 0.52 & 0.37 \\
\enddata
\tablecomments{
A portion of the table is shown for guidance regarding its format.
The entire table is available in machine-readable form in the online journal. }
\tablenotetext{a}{Richness, photometric redshift of redMaPPer clusters and the number of cluster members with $P_{mem} > 0$ given in \citet{Rykoff16}. }
\tablenotetext{b}{Spectroscopically determined redshift and the number of spectroscopically identified members derived in this study. }
\tablenotetext{c}{Spectroscopic completeness for redMaPPer clusters (equation \ref{eq_frac1}). }
\tablenotetext{d}{Spectroscopically identified member fraction in redMaPPer clusters (equation \ref{eq_frac2}). }
\label{hmapred}
\end{deluxetable*} 

\begin{deluxetable*}{lccccccc}
\tablecolumns{8}
\tabletypesize{\scriptsize}
\tablewidth{0pt}
\tablecaption{HectoMAP-red clusters}
\tablehead{
\colhead{Cluster ID} & \colhead{SDSS Object ID} & \colhead{R.A.}     & \colhead{Decl.} & 
\colhead{$P_{mem}$}  & \colhead{$z_{spec}$}     & \colhead{z Source} & \colhead{Spec. Mem}}
\startdata
05570 & 1237661361301684348 & 213.523263 &  43.472578 & 0.6437 & $0.08888 \pm 0.00002$ & SDSS & N \\
05570 & 1237661434317242581 & 213.573735 &  43.599875 & 0.2995 & $0.08996 \pm 0.00002$ & SDSS & N \\
05570 & 1237661434317242611 & 213.661584 &  43.639202 & 0.5661 & $0.09040 \pm 0.00007$ & MMT  & N \\
05570 & 1237661361301684354 & 213.560132 &  43.530140 & 0.8786 & $0.08956 \pm 0.00005$ & MMT  & N \\
05570 & 1237661434317242524 & 213.581176 &  43.749526 & 0.7643 & $0.11388 \pm 0.00002$ & SDSS & N \\
05570 & 1237661434317176918 & 213.354990 &  43.759024 & 0.7093 & $0.08890 \pm 0.00003$ & SDSS & N \\
05570 & 1237661434317176951 & 213.420309 &  43.761725 & 0.9106 & $0.09206 \pm 0.00002$ & SDSS & N \\
05570 & 1237661434317176965 & 213.411436 &  43.709942 & 0.9576 & $0.08951 \pm 0.00002$ & SDSS & N \\
05570 & 1237661434317176989 & 213.423029 &  43.682213 & 0.9559 & $0.08877 \pm 0.00003$ & SDSS & N \\
05570 & 1237661434317176998 & 213.444344 &  43.714732 & 0.9337 & $0.08559 \pm 0.00002$ & SDSS & N
\enddata
\tablecomments{
A portion of the table is shown for guidance regarding its format.
The entire table is available in machine-readable form in the online journal. }
\label{hmapredmem}
\end{deluxetable*}  

The HectoMAP census of the 104 redMaPPer cluster candidates 
 includes 2641 cluster members defined by straightforward cuts in redshift space. 
These members yield a mean redshift for each cluster and 
 a spectroscopic measure of the richness that we investigate further in Section \ref{richness}. 
The spectroscopy underscores several issues in photometric cluster identification
 including the capture of galaxies in foreground and background structures as potential cluster members
 and mismatch between the spectroscopically identified BCG and the redMaPPer central galaxies ($\sim 25\%$ of the time). 

\begin{figure*}
\centering
\includegraphics[scale=0.65]{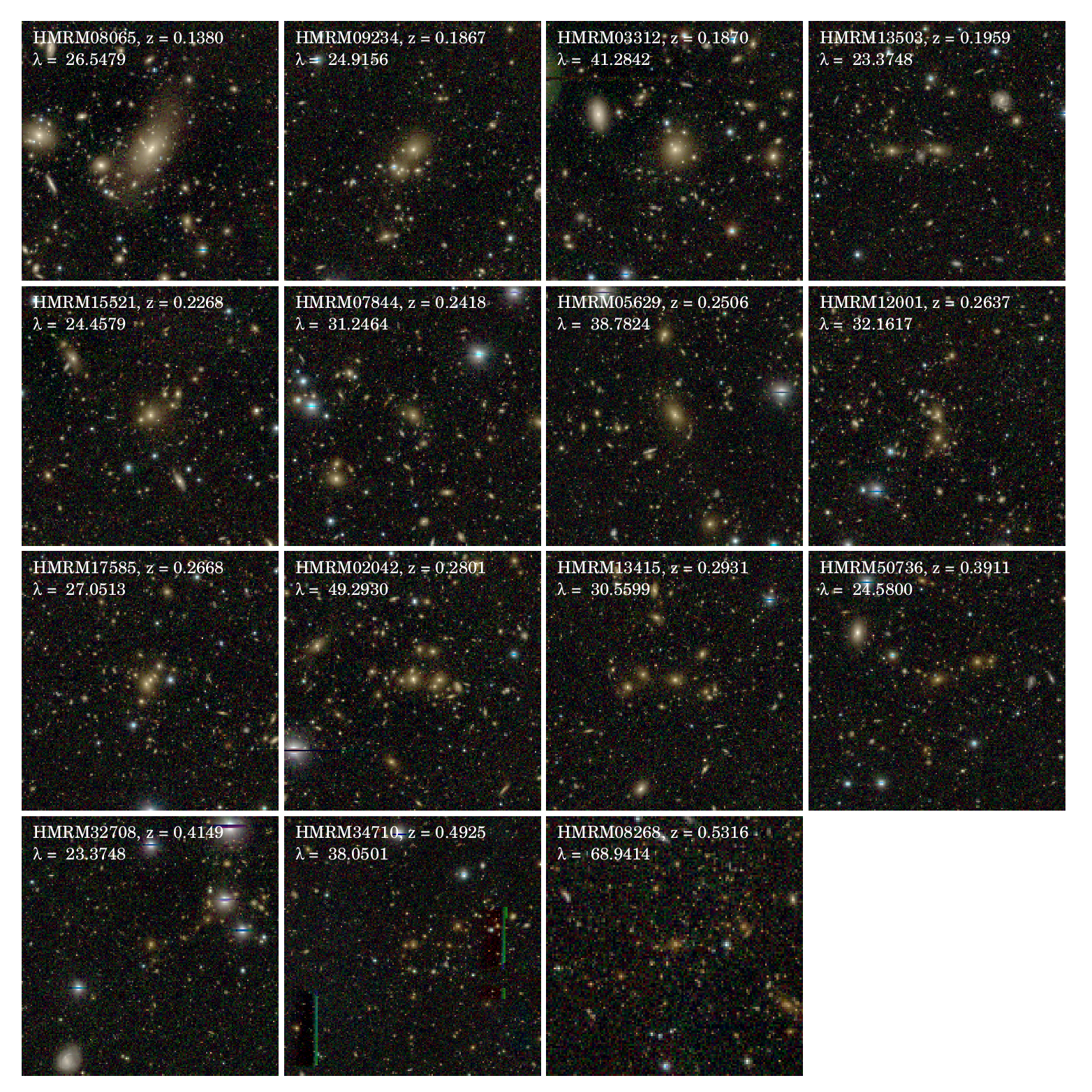}
\caption{{\it Subaru}/HSC thumbnail images of HectoMAP redMaPPer clusters in the HSC public archive
 sorted by their redshifts. 
The image sizes are 2 arcmin $\times$ 2 arcmin, except HMRM08268 (1.5 arcmin $\times$ 1.5 arcmin).
The color channel R,G,B of the thumbnails are HSC-i, HSC-r, HSC-G, respectively.}
\label{hsc}
\end{figure*}

\subsection {{\it Subaru}/HSC SSP Imaging}

Fifteen systems lie within the the HSC Subaru Strategic Program DR1 \citep{Aihara17},
 which covers 7 deg$^{2}$ ($\sim 13\%$) of the HectoMAP region. 
Figure \ref{hsc} shows the HSC images of the individual systems.
Each thumbnail image displays a $2\arcmin \times 2\arcmin$ field of view 
 around the redMaPPer center, except for HMRM08268 at $z = 0.5133$.
Because HMRM08268 is at the edge of the public archive, 
 we show only a $1.5\arcmin \times 1.5\arcmin$ field of view for this cluster.
 
The HSC images demonstrate that 
 most HectoMAP-red clusters are obvious rich clusters.
In a few cases including the systems (HMRM13503, HMRM32708, HMRM34710) 
 there are only a few red objects in the image possibly suggesting that 
 the system is a poor group rather than a rich cluster. 
We note that HMRM13503 at $z_{phot} = 0.1959$ and HMRM32708 at $z_{phot} = 0.4122$
 have very low redMaPPer richness $\lambda_{rich} \sim 23$. 
For HMRM34710 at $z_{phot} = 0.4933$, 
 the richness ($\lambda_{rich} =38.05$) appears to be overestimated 
 based both on the {\it Subaru}/HSC image and on the HectoMAP spectroscopy. 

The imaging suggests that 3/15 of these redMaPPer candidates may not be rich clusters.  
One would expect that in these deep images the faint cluster population, 
 invisible to the SDSS limiting magnitude, would be apparent. 
We comment further on the spectroscopy of these systems in Section \ref{spec}.

\subsection{HectoMAP spectroscopy and redMaPPer Members}\label{spec}

\begin{figure}
\centering
\includegraphics[scale=0.49]{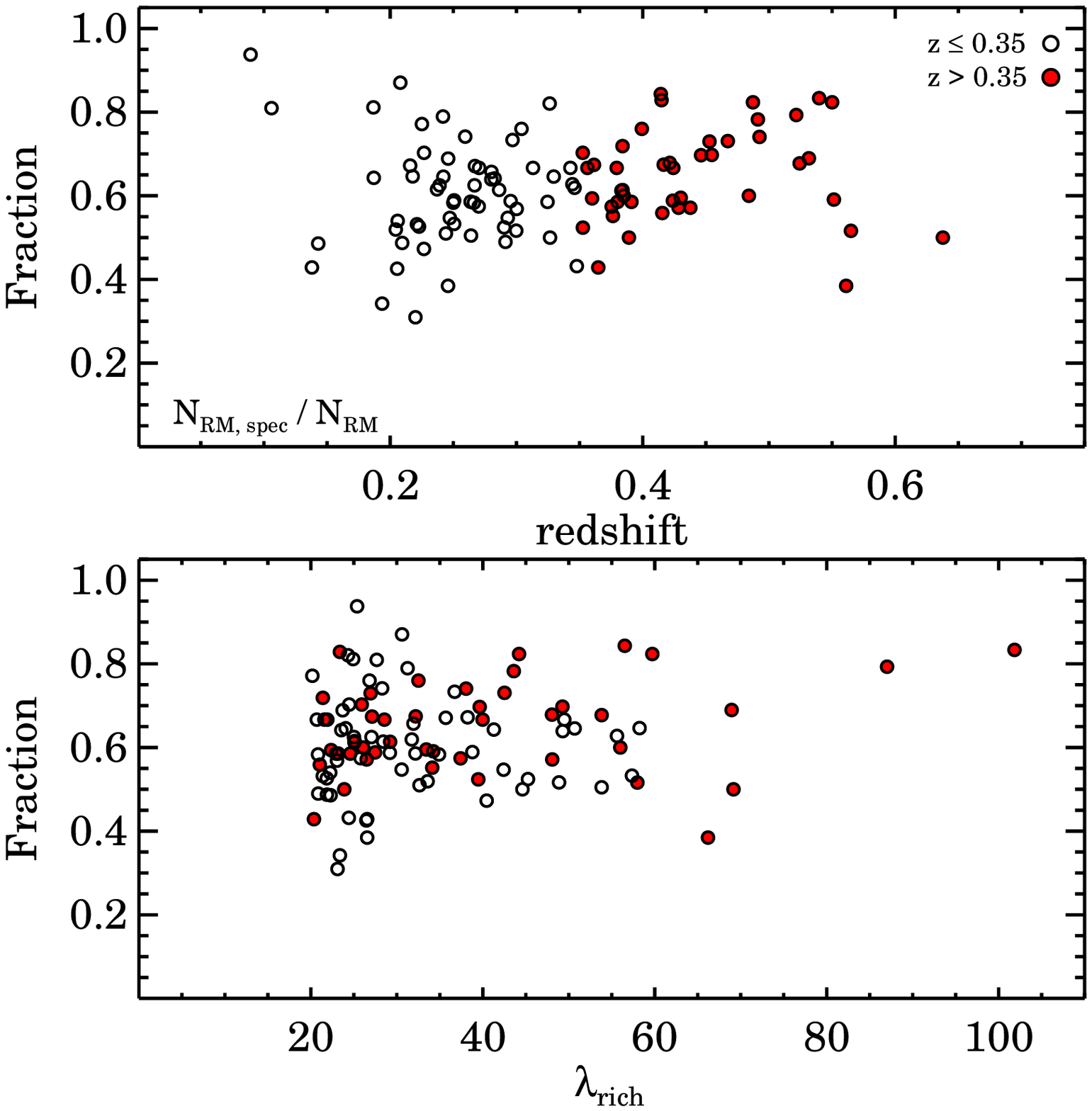}
\caption{Spectroscopic completeness of redMaPPer members ($P_{mem} > 0$) in HectoMAP
 as a function of redshift (upper panel) and richness, $\lambda_{rich}$ (lower panel). }
\label{clcomp}
\end{figure}

As a first step in evaluating the redMaPPer candidate clusters, 
 we measure the spectroscopic completeness for individual HectoMAP-red clusters 
 as a function of redshift and richness (Figure \ref{clcomp}). 
We define the spectroscopic completeness as 
\begin{equation}
f_{comp} = \frac{N_{RM, spec}}{N_{RM}}, 
\label{eq_frac1}
\end{equation}
 where $N_{RM, spec}$ is the number of redMaPPer member candidates 
 (redMaPPer membership probability $P_{mem} > 0$) with spectra, and 
 $N_{RM}$ is the total number of redMaPPer member candidates brighter than $r_{petro, 0} = 21.3$, 
 the limiting apparent magnitude of HectoMAP. 

For $\sim 90\%$ of the clusters, 
 HectoMAP includes redshifts for $\gtrsim 50\%$ of the member candidates.
There are no strong trends of spectroscopic completeness with redshift or richness. 
This independence results from the $r = 21.3$ limit of HectoMAP 
 that is reasonably close to the limiting $r \sim 22$ used by redMaPPer. 
Thus, measurement of the spectroscopic properties of the HectoMAP-red clusters 
 should be insensitive to any HectoMAP sampling biases. 
 
With the HectoMAP sample for each HectoMAP-red cluster, 
 we identify spectroscopic members and revise the cluster mean redshift. 
We identify cluster members in the phase space defined by 
 the rest-frame relative velocity difference as a function of clustercentric distance, the R-v diagram.
In the R-v diagram, 
 the cluster members show a strong concentration around the cluster center 
 (e.g. \citealp{Diaferio97, Rines06, Rines13, Rines16, Serra13, Sohn17a}). 
 
Following previous studies, 
 we identify cluster members based on the R-v diagrams. 
Here we apply a simple boundary
 because many systems are not very well populated: 
 $R_{cl} < 1.5$ Mpc and $|\Delta c (z_{galaxy} - z_{cl}) / (1 + z_{cl})| < 2000~\kms$,
 where $R_{cl}$ is the clustercentric distance, $z_{galaxy}$ is the spectroscopic redshift of galaxies in the field, 
 $z_{cl}$ is the cluster central redshift. 
We set the $R_{cl}$ limit based on the maximum $R_{cl}$ of redMaPPer members with $P_{mem} > 0$ and 
 the $|\Delta c (z_{galaxy} - z_{cl}) / (1 + z_{cl})|$ limit 
 based on the maximum range of 
 spectroscopically identified members in known massive clusters (e.g. HeCS, \citealp{Rines13, Rines16}).
The redMaPPer spectroscopic membership is insensitive to the redshift cut from $\sim 1500 - 2500~\kms$. This cut is 60\% or less of the photometric redshift window.
The spatial and redshift limits are necessarily generous compared with techniques applicable to better sampled systems 
 (e.g. \citealp{Rines06, Rines16, Sohn17a}). 

To identify cluster members, 
 we examine the R-v diagrams based on the cluster center from the original redMaPPer cluster catalog \citep{Rykoff16}. 
For the cluster central redshift, 
 we first check the redshift of the central galaxy identified by redMaPPer, $z_{central}$.
The HectoMAP redshift survey includes 
 spectroscopic redshifts of 100 (96\%) of the HectoMAP-red central galaxies. 
If there is a redshift for the central galaxy, 
 we identify cluster members by applying the $R_{cl}$ and $|\Delta c (z_{galaxy} - z_{central}) / (1 + z_{central})|$ window. 
For some clusters, including the four systems without a spectroscopic redshift of the central galaxy, 
 there are still only a few spectroscopically confirmed redMaPPer members around the central galaxy. 

We estimate 
 the median spectroscopic redshift, $z_{med}$, of the redMaPPer members with $P_{mem} > 0$. 
We re-identify the spectroscopic members from the R-v diagrams centered on this revised $z_{med}$. 
Finally, we take either $z_{central}$ or $z_{med}$ as the estimate of the cluster mean. 
We choose the estimate based on the largest number of plausible spectroscopic members. 
Hereafter, the cluster redshift ($z_{spec, cl}$) is
 the central redshift of a HectoMAP-red cluster 
 determined from spectroscopically identified members.   

\begin{figure}
\centering
\includegraphics[scale=0.49]{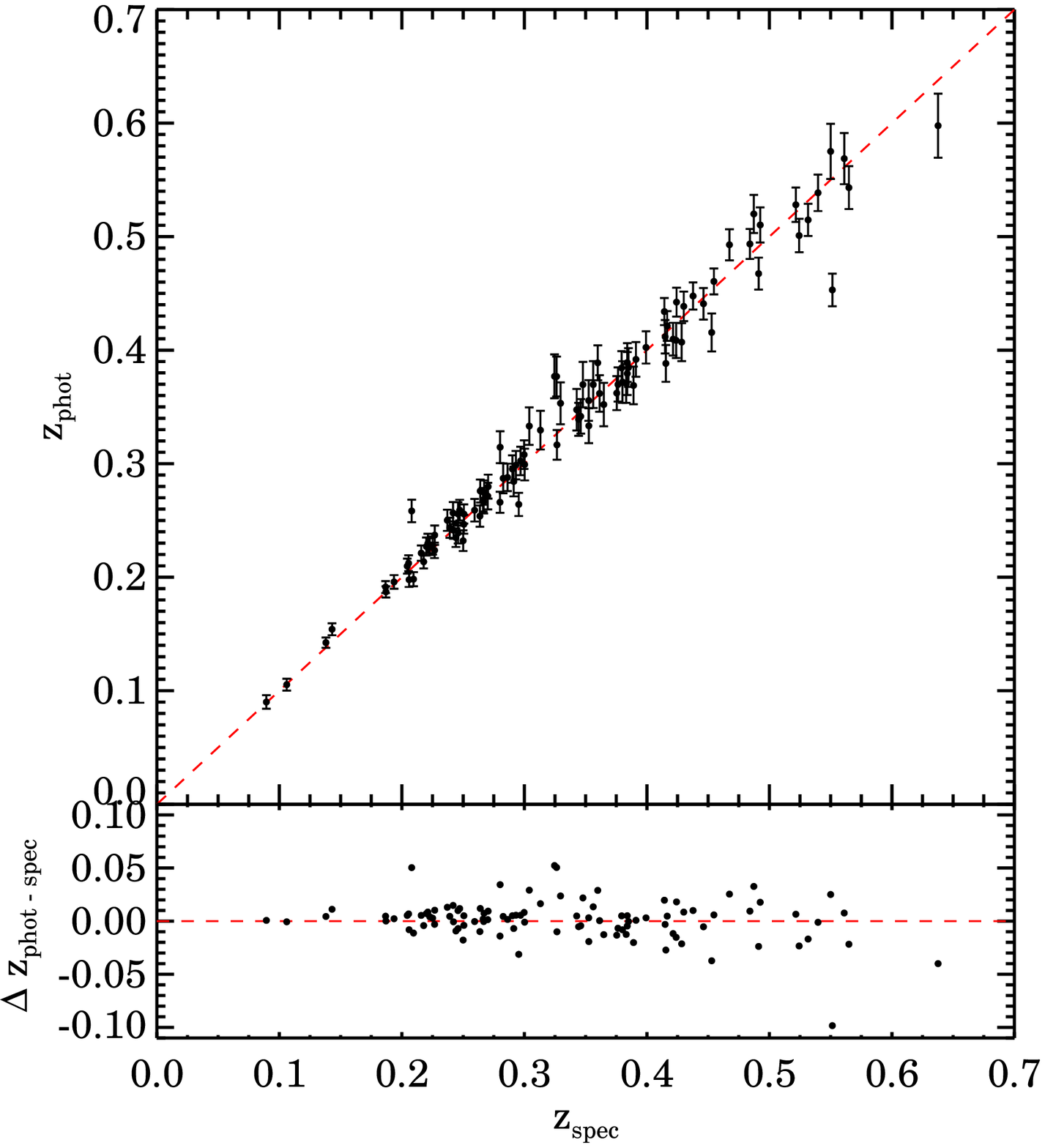}
\caption{(Upper) Photometric redshifts vs. spectroscopic redshifts of HectoMAP redMaPPer clusters and 
(Lower) the difference between them. 
Here spectroscopic redshift indicates the mean spectroscopic redshift of cluster members (if any). }
\label{zcomp}
\end{figure}

Figure \ref{zcomp} compares the spectroscopic ($z_{spec,cl}$) and photometric redshifts ($z_{phot,cl}$) of 
 all of the HectoMAP-red clusters. 
The photometric redshifts are generally consistent with the spectroscopic redshifts with a few significant offsets. 
The mean difference ($|\Delta c (z_{phot,cl} - z_{spec, cl}) / (1 + z_{spec, cl})|$) is $\sim 3800$ km s$^{-1}$,  
 comparable with the mean cluster photometric redshift uncertainty for a single cluster in the redMaPPer catalog 
 (i.e. $\sim 3800$ km s$^{-1}$). 
It is noteworthy that 
 the $3 \sigma$ photometric error is comparable with 
 the diameter of smaller voids in the HectoMAP survey (see Section \ref{comp}). 
For the 11 systems with $|\Delta c (z_{phot, cl} - z_{spec, cl}) / (1 + z_{spec, cl})| > 6000~\kms$, 
 our spectroscopic survey is relatively incomplete ($< 50\%$). 
The redMaPPer catalog identified $\sim 48-75$ members for these 11 systems, 
 but we identify only $\sim 6-35$ spectroscopic members. 
 
\begin{figure*}
\centering
\includegraphics[scale=0.65]{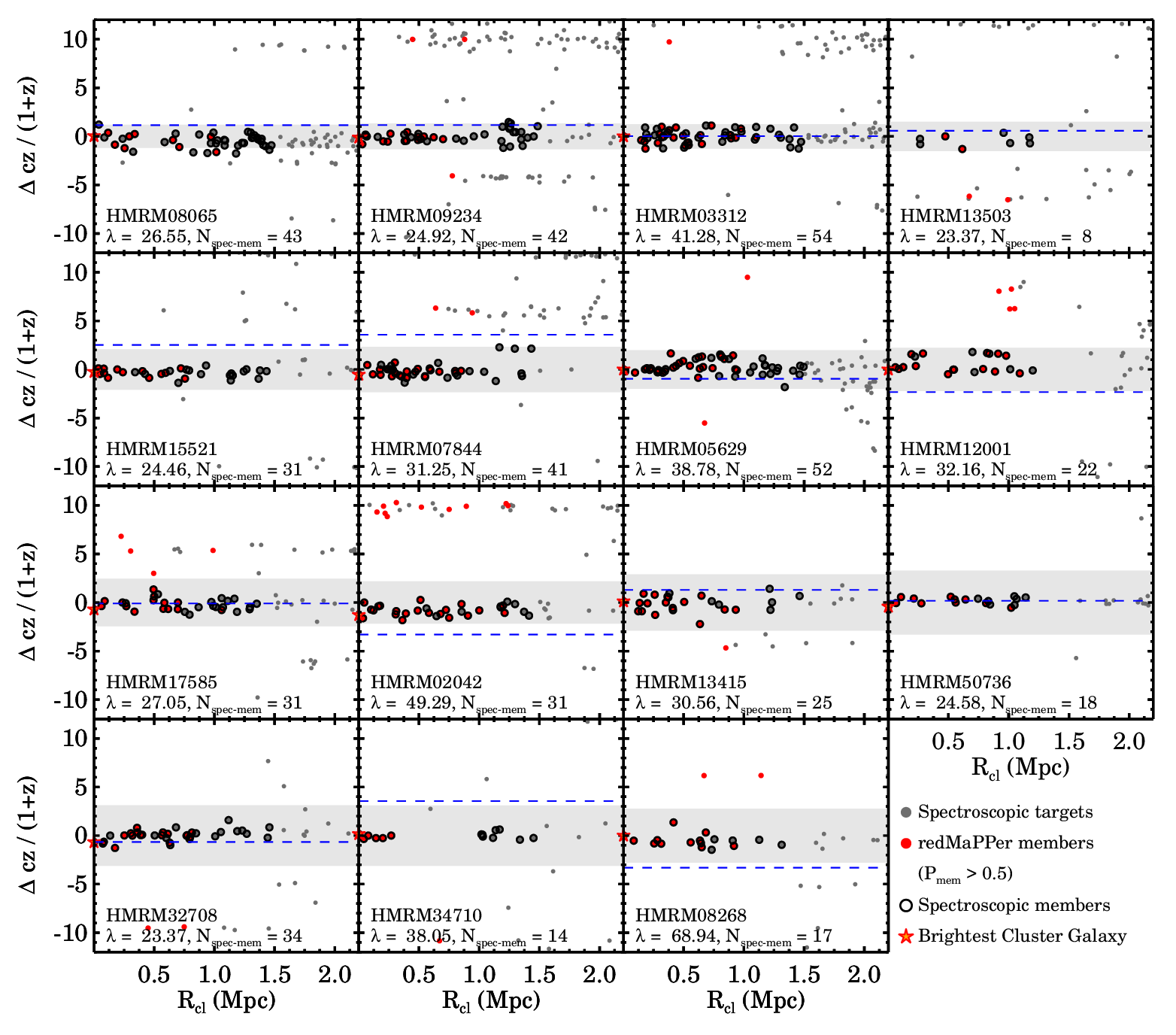}
\caption{Rest-frame clustercentric radial velocities vs. clustercentric radius 
 for HectoMAP redMaPPer clusters shown in Figure \ref{hsc}. 
Here, the spatial centers are from the redMaPPer catalog, and 
 the central redshifts are spectroscopically determined. 
Gray dots are spectroscopic targets within the fields.
Red filled circles and black open circles are redMaPPer members with $P_{mem} > 0.5$ and 
 the spectroscopically identified members, respectively. 
Starlets indicate the brightest cluster galaxies. 
Dashed lines mark the cluster photometric redshift given in the redMaPPer catalog. 
Gray shaded regions show the uncertainty of the photometric redshift of the HectoMAP-red clusters
 from \citet{Rykoff16}. 
The $\Delta cz / (1 + z_{cl})$ cuts for identifying spectroscopic members we use
 corresponds to $\sim 60\%$ of the photometric redshift uncertainty of the clusters. }
\label{rv}
\end{figure*}
 
Figure \ref{rv} shows the R-v diagrams of 
 the 15 HectoMAP-red clusters with HSC images. 
We use the redMaPPer center 
 and the spectroscopically determined cluster redshift. 
 
Figure \ref{rv} includes the R-v diagrams for the three systems 
 that are not apparent rich clusters in the HSC images. 
HMRM13503 ($z_{phot} = 0.1959$) contains only eight spectroscopic members within the membership window. 
The redMaPPer algorithm identifies more members in the field, 
 but these redMaPPer members are foreground and background objects. 
HMRM32708 ($z_{phot} = 0.4122$) has 34 spectroscopic members, but the redMaPPer richness is quite low. 
There are 14 spectroscopic members in the HMRM34710 ($z_{phot} = 0.4933$) field. 
However, only six members are located around the BCG
 and eight members are well separated from the BCG.  
Thus, HMRM13503 and HMRM34710 are poor groups
 as both the HSC imaging and the R-v diagrams suggest. 
 
In 76 of the 104 clusters, 
 the spectroscopic brightest cluster galaxy (BCG) is identical to the redMaPPer central galaxy. 
For four systems, 
 we lack a spectroscopic redshift for the redMaPPer central galaxy. 
However, in 24 systems ($\sim 23\%$), 
 the spectroscopy identifies a BCG which is not the redMaPPer central galaxy. 
This fraction of offset BCGs 
 is comparable with the redMaPPer estimate of the number of probable central galaxy misidentifications (15-20\%, \citealp{Rykoff16}). 
 
In Figure \ref{rv}, gray dots mark the spectroscopic targets and 
 red filled circles indicate HectoMAP-red member candidates with $P_{mem} > 0.5$. 
The dashed lines indicate the photometric redshifts of the clusters assigned by redMaPPer. 
Note that the photometric redshifts are often significantly offset 
 from the mean spectroscopic redshift of the HectoMAP-red member candidates (red filled circles). 

redMaPPer members generally cluster around the BCG. 
Interestingly, some redMaPPer members are not at the cluster redshift.
These redMaPPer members tend to have low membership probability (see Section \ref{cont}). 
Furthermore,
 a significant fraction of the spectroscopically identified members are not 
 identified by the redMaPPer algorithm. 
The objects redMaPPer fails to evaluate are a mix of galaxies much bluer than the red sequence and of 
 apparent failures of the redMaPPer algorithm to identify true red cluster members.
 
\begin{figure}
\centering
\includegraphics[scale=0.49]{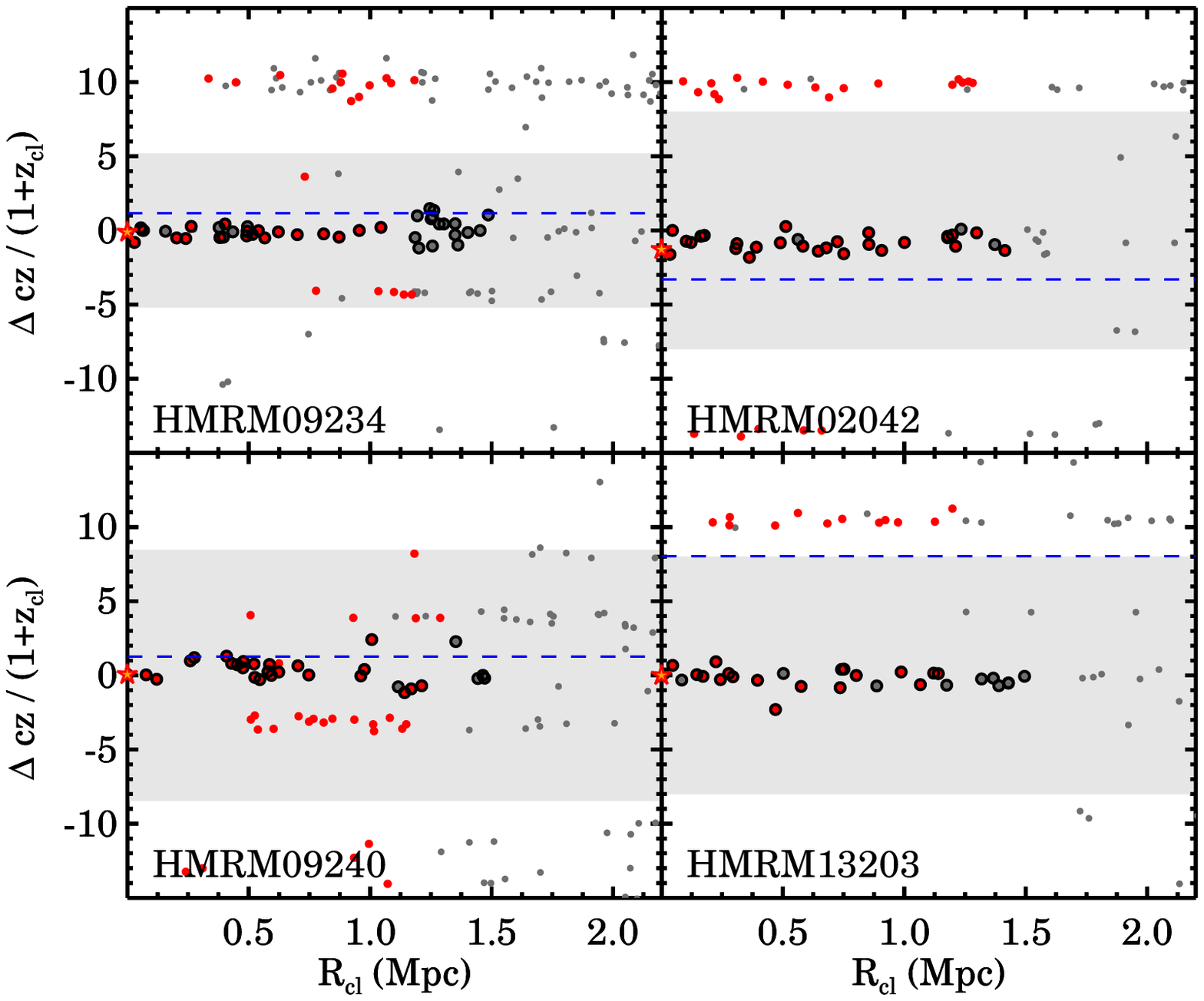}
\caption{
R-v diagrams of example HectoMAP-red clusters with dense superimposed structure along the line-of-sight. 
The symbols are the same as Figure \ref{rv}
 except for the red filled circles and the gray shaded regions. 
The red filled circles show the redMaPPer members with $P_{mem} > 0$. 
The shaded region displays the uncertainty in the photometric redshift error 
 for individual galaxies (Figure 7 of \citealp{Rykoff14}). 
Note that foreground/background structures 
 contain a number of galaxies comparable with the main cluster.} 
\label{rv_ovl}
\end{figure}
 
The R-v diagrams show 
 some of the neighboring foreground and background structures 
 in the line-of-sight direction toward the HectoMAP-red clusters. 
Figure \ref{rv_ovl} shows R-v diagrams of four HectoMAP-red clusters
 where there are dense structures within the photometric redshift window for individual galaxies
 (indicated by gray shading). 
In these cases,
 the redMaPPer member candidate list includes large numbers of these foreground and background objects 
 (red dots). 
This inclusion of nearby structures artificially inflates 
 the redMaPPer richness of these systems.

\subsection {The Red Sequence of HectoMAP-red Clusters}

\begin{figure*}
\centering
\includegraphics[scale=0.65]{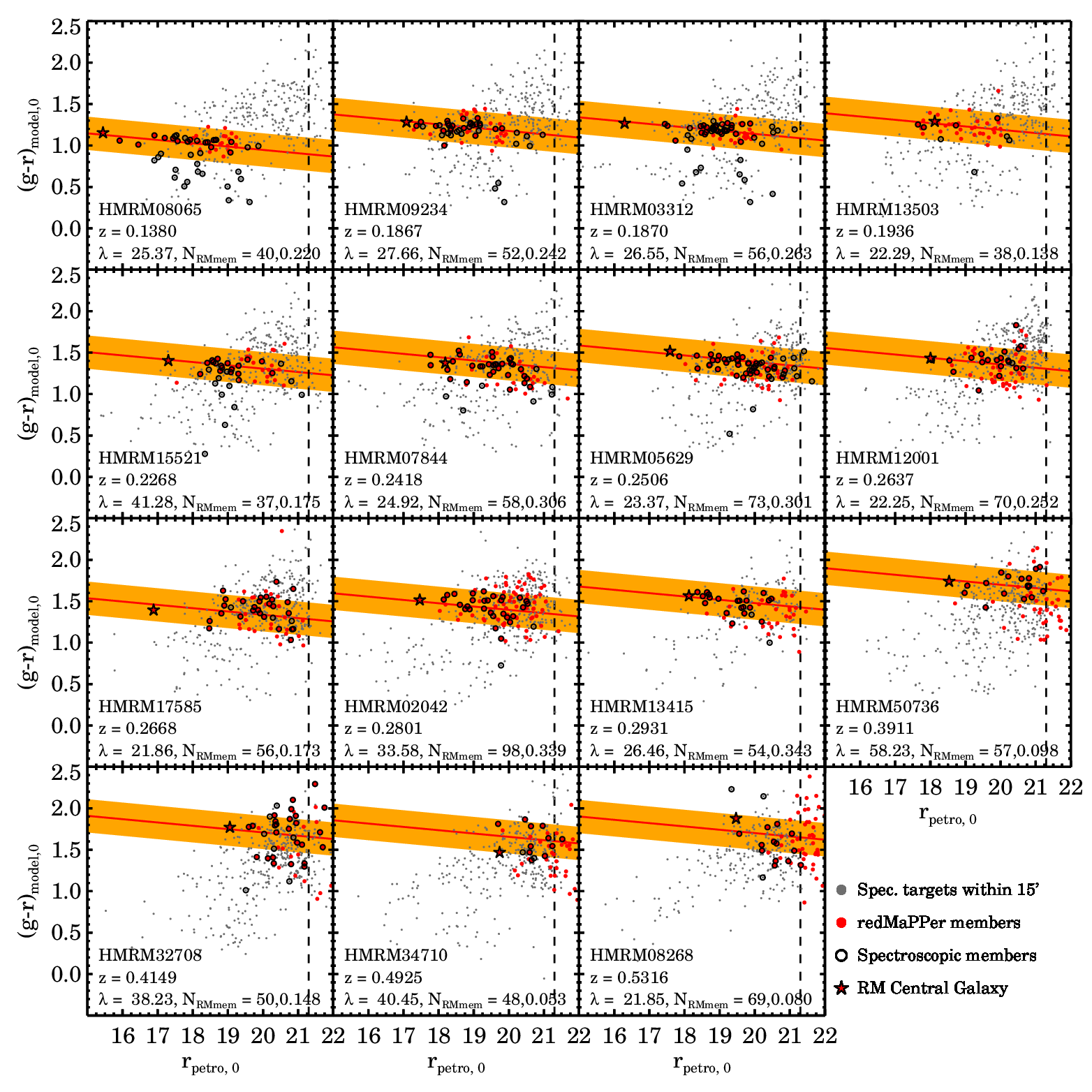}
\caption{$(g-r)_{model, 0}$ vs. $r_{petro, 0}$ color-magnitude diagrams of 
 HectoMAP-red clusters shown in Figure \ref{hsc}. 
Gray dots are spectroscopic targets within $15\arcmin$ of the cluster center and 
 red filled circles are redMaPPer members with $P_{mem} > 0$. 
Black open circles indicate spectroscopically identified members.
Shaded regions display the red-sequence (solid line) $\pm 0.2$.}
\label{cmd}
\end{figure*}

The redMaPPer algorithm identifies clusters on the basis of the red sequence. 
To explore the prominence of the red sequence, 
 Figure \ref{cmd} shows the $(g-r)_{model, 0}$ versus $r_{petro, 0}$ color magnitude diagrams 
 of the 15 HectoMAP-red clusters with HSC images. 
We plot galaxies within 15 arcmin of each cluster center
 as gray dots. 
The red filled circles and black open circles indicate 
 redMaPPer members and the full set of spectroscopically identified members, respectively. 
Following \citet{Rines13}, 
 we determine the $g-r$ red-sequence of each cluster
 by assuming a slope of $-0.04$ in the color-magnitude domain and
 fitting the spectroscopically identified member galaxies; 
 we then identify red-sequence members as objects within $\pm~0.2$ mag of the red-sequence. 
The slope of the red-sequence may change at higher redshift,
 but galaxies in the well-populated HectoMAP X-ray clusters ($z \lesssim 0.4$, \citealp{Sohn17c})
 are consistent with this definition of the red sequence. 
  
The red-sequence we use differs from the one used by redMaPPer. 
We determine the red-sequence in the $(g - r)_{model, 0} - r_{cmodel, 0}$ space; 
 redMaPPer derives the multicolor red-sequence model based on all of the SDSS bands. 
The redMaPPer algorithm also derives 
 the slope of the red-sequence model for each cluster rather than assuming a single value.  
Here, we use the red sequence only to show the impact of spectroscopy on  redMaPPer cluster identification. 
We do not use the red sequence itself to identify real systems. 
However, for 11 of the 15 clusters the $g-r$ versus $r$ red sequence is well-defined. 
When this red sequence is not readily visible, 
 the richness of the system appears to be substantially overestimated by redMaPPer and/or
 the system is at $z \geq 0.4$ where the scatter in the red sequence is substantial in this color-magnitude space. 

We examine the $(r-i)_{model, 0} - i_{petro, 0}$ color-magnitude diagrams for clusters with $z \geq 0.35$.
For the clusters at $z > 0.35$, 
 we determine the $r-i$ red-sequence for each cluster by assuming a slope -0.01 and 
 we select red-sequence members within $\pm0.1$ of the red-sequence. 
This choice of slope provides a reasonable representation of the data.
In general, the red-sequences of clusters with $z \geq 0.35$ are flatter and tighter 
 in the $r-i$ color domain (e.g. Figure 3 from \citealp{Rykoff14}). 
The scatter around the red sequence increases with redshift (e.g. Figure 4 from \citealp{Rykoff14}). 

In the color-magnitude diagram for each cluster, 
 a significant number of non-members (generally background objects) contaminate 
 the apparent red sequence and bias the richness estimate upward. 
There are also spectroscopically determined members 
 that lie on the approximate red sequence we define,
 but they do not have a redMaPPer membership probability. 
This problem may originate from large offsets
 between the photometric redshift reported by redMaPPer and 
 the more accurate spectroscopic mean redshift. 
 
\section{DISCUSSION}\label{disc}

HectoMAP enables a direct examination of the spectroscopic properties of the  
 104 photometrically selected redMaPPer clusters (HectoMAP-red) covering the redshift range $0.08 < z < 0.6$. 
Because HectoMAP targets red galaxies, 
 the redshift survey is particularly powerful for investigating 
 clusters identified with a red-sequence technique like the one applied to identify redMaPPer candidate systems.  
HectoMAP includes redshifts for $\gtrsim 60\%$ of the redMaPPer cluster candidate members with $P_{mem} > 0$. 
 
Photometric cluster selection is obviously subject to 
 contamination by unrelated structures along the line-of-sight. 
We examine the frequency of these line-of-sight structures 
 in the 104 HectoMAP-red clusters (Section \ref{cont}).  
In Section \ref{richness}, 
 we explore the spectroscopically determined richness relative to the redMaPPer richness of these systems.  
In Section \ref{comp}, 
 we discuss the fraction of HectoMAP-red clusters 
 that are confirmed with spectroscopy and 
 we discuss indications that the redMaPPer catalog is not a complete census even of the richest clusters in the HectoMAP region.

\subsection{Superpositions along the Line-of-Sight}\label{cont}

\begin{figure}
\centering
\includegraphics[scale=0.49]{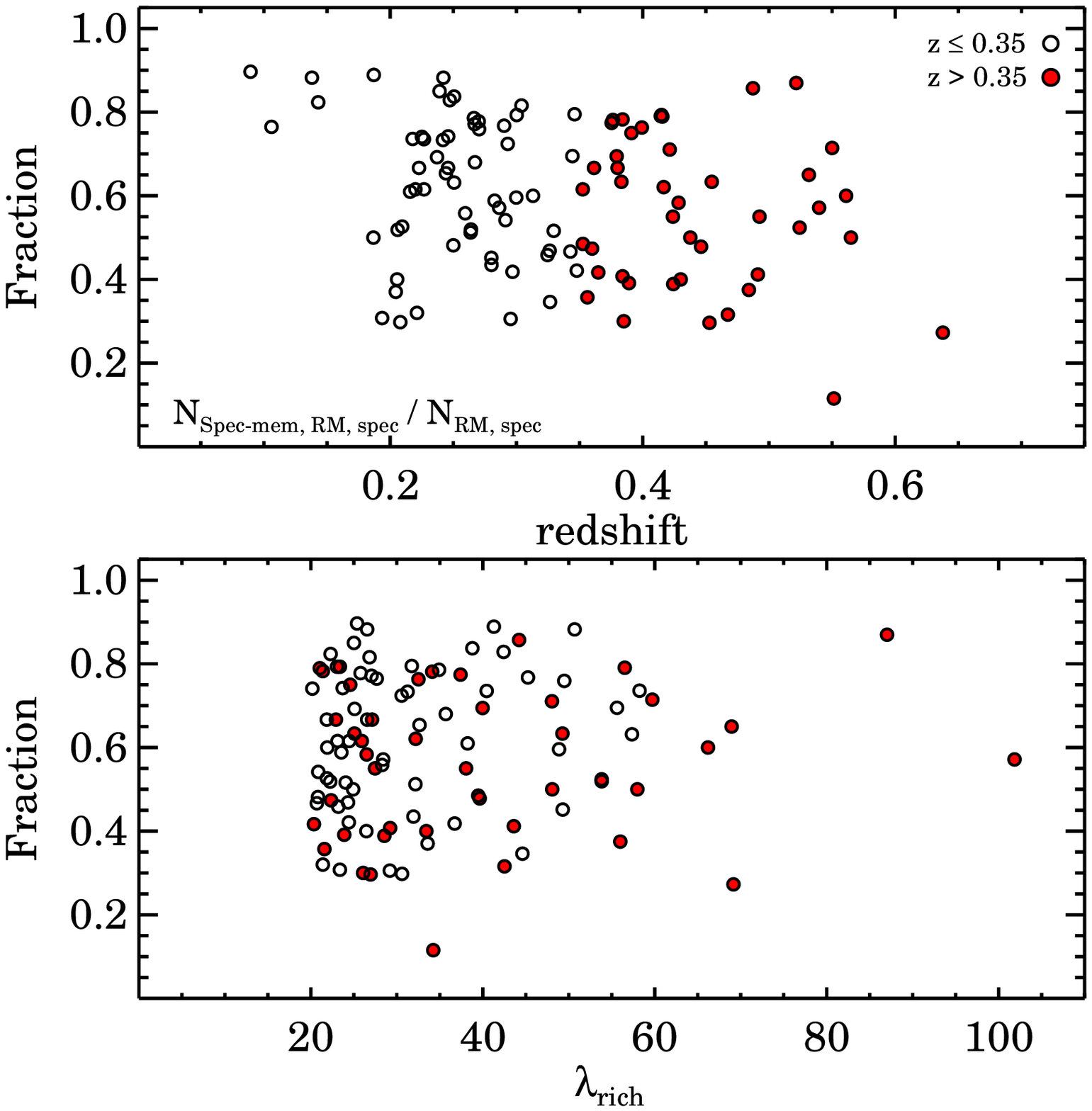}
\caption{
The fraction of spectroscopically identified members 
 among redMaPPer members ($P_{mem} > 0$) with spectroscopic redshifts
 as a function of cluster redshift (upper) and richness ($\lambda_{rich}$, lower). 
More than 20\% of redMaPPer members are contaminant foreground or background objects. }
\label{frac2}
\end{figure}

Spectroscopic surveys of galaxy clusters 
 allow estimation of the contamination of candidate members selected by color. 
Figure \ref{frac2} shows the spectroscopically identified member fraction among the redMaPPer candidate members.
We define the spectroscopically identified member fraction in each cluster as 
\begin{equation}
f_{spec-mem, cl} = \frac{N_{spec-mem, RM, spec}}{N_{RM, spec}},
\label{eq_frac2}
\end{equation}
 where $N_{spec-mem, RM, spec}$ is the number of spectroscopically identified members 
 among the redMaPPer candidate members and 
 $N_{RM, spec}$ is the total number of redMaPPer candidate members with $P_{mem} > 0$ and with spectra. 
The median $f_{spec-mem, cl}$ of the HectoMAP-red clusters is $\sim 59\%$. 
The median $f_{spec-mem, cl}$ increases 
 for candidate members with 
 $P_{mem} > 0.5$ ($\sim 72\%$) and $P_{mem} > 0.9$ ($\sim 91\%$). 
We find no dependence of the spectroscopically identified member fraction on redshift and richness. 

\begin{figure*}
\centering
\includegraphics[scale=0.65]{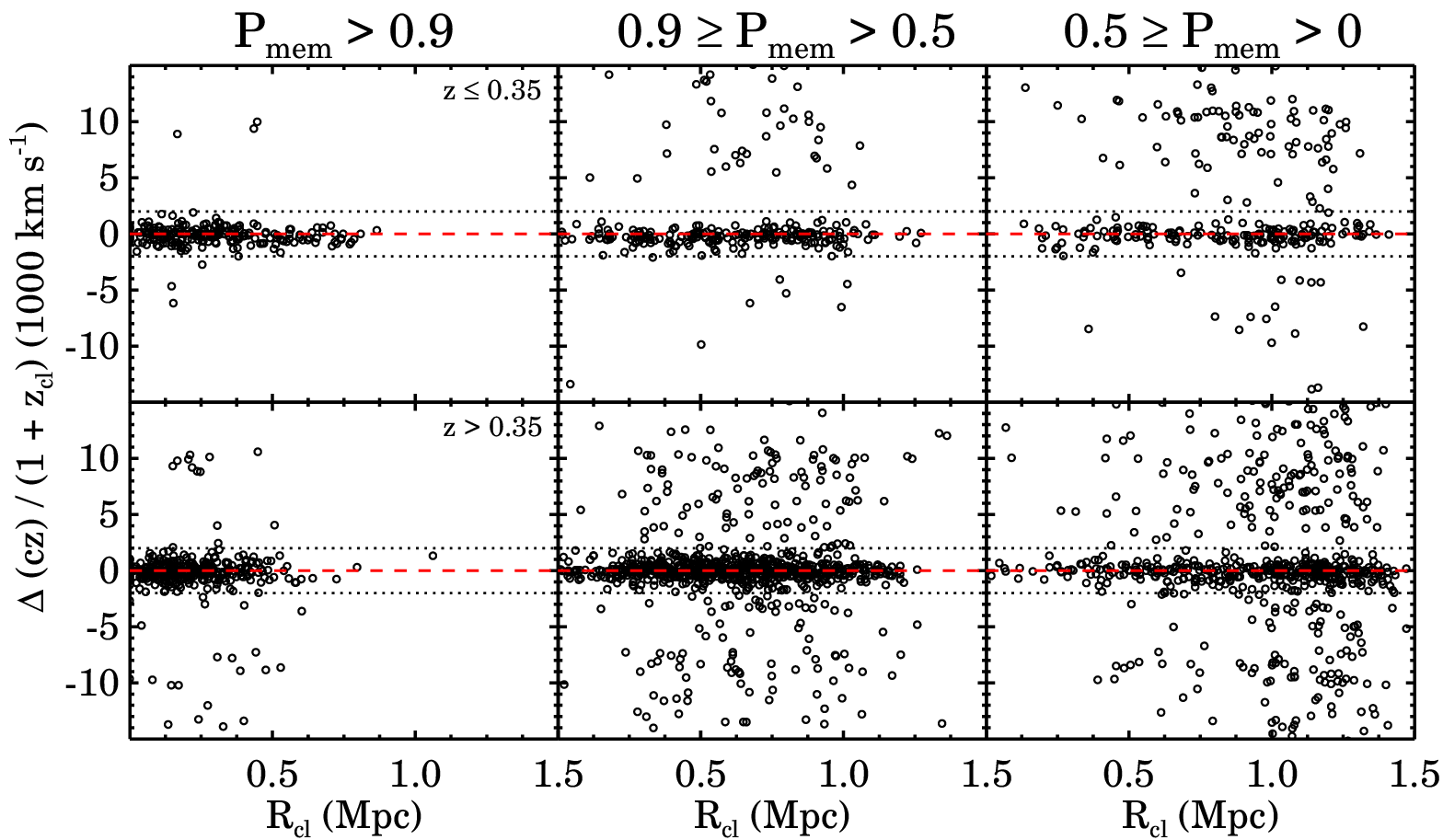}
\caption{The stacked R-v diagrams of 104 HectoMAP-red clusters
 for three different bins of membership probability:
 $P_{mem} \geq 0.9$, $0.9 > P_{mem} > 0.5$, $0.5 > P_{mem}$
 for clusters with $z \leq 0.35$ (upper) and with $z > 0.35$ (lower).
The dotted lines show the redshift window we use for identifying spectroscopic members.  
Note that there are many galaxies with $P_{mem} > 0.5$ 
 offset from the cluster center along the line-of-sight. }
\label{rv_pmem}
\end{figure*}

The R-v diagrams in Figure \ref{rv_pmem} show how
 the distribution of the spectroscopic HectoMAP-red cluster members  
 depends on $P_{mem}$.
We plot stacked R-v diagrams of the 
 61 low redshift ($z \leq 0.35$) and 43 high redshift ($z > 0.35$) HectoMAP-red clusters 
 within three different $P_{mem}$ bins:
 $P_{mem} \geq 0.9$, $0.9 > P_{mem} \geq 0.5$, and $0.5 > P_{mem}$. 
For $P_{mem} \geq 0.9$,
 there is a strong concentration of spectroscopically identified members toward the cluster center. 
Even for these high confidence objects, 
 $\sim11\%$ have a rest-frame relative velocity that 
 differs from the cluster center by ($|\Delta cz / (1+z_{cl})| \geq 2000~\kms$.
The typical velocity dispersion derived from the stacked $P_{mem} \geq 0.9$ members
 is $\sim 650~\kms$ comparable with the typical rich cluster 
 line-of-sight velocity dispersion ($\sim 700~\kms$, \citealp{Rines13}). 
The candidate members with lower $P_{mem}$ generally lie 
 at larger radius if they are within the redshift range for spectroscopic membership. 
The fraction of objects with large $|\Delta (cz) / (1+z_{cl})|$ increases as $P_{mem}$ decreases:
 $\sim35\%$ are outliers when $0.9 > P_{mem} \geq 0.5$ and 
 $\sim66\%$ outliers are when $0.5 > P_{mem}$.

Based on the fraction of spectroscopically identified members 
 among redMaPPer member candidates, 
 we provide a correction function for redMaPPer cluster membership. 
\citet{Rozo15b} also estimate the spectroscopic member fraction 
 among redMaPPer members, 
 but they use only ‘red’ galaxies from various spectroscopic samples.

\begin{figure}
\centering
\includegraphics[scale=0.49]{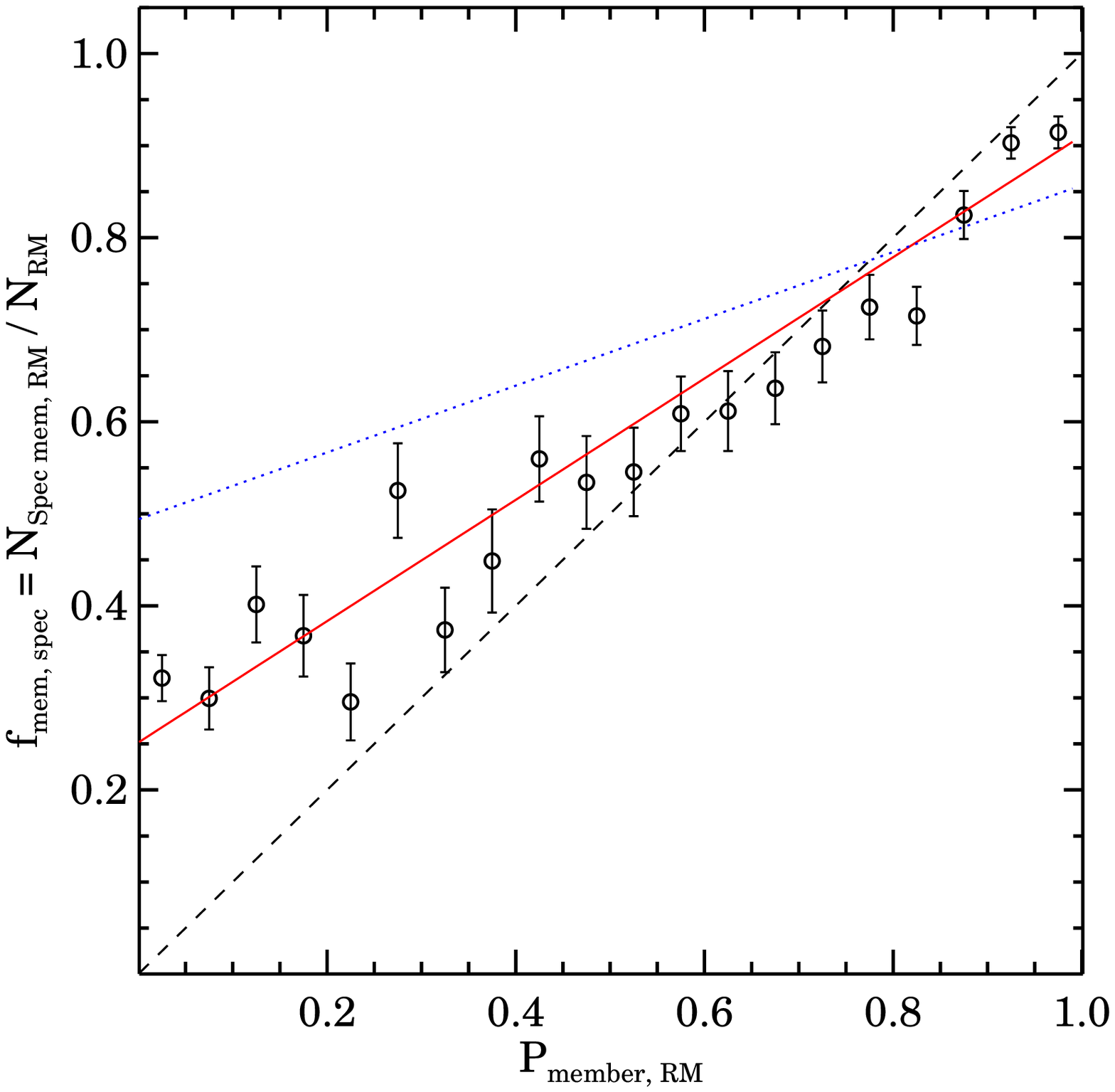}
\caption{The fraction of spectroscopically identified members among redMaPPer members with spectroscopic redshifts
 as a function of redMaPPer membership probability ($P_{mem}$). 
Error bars indicate the standard deviation for 1000 bootstrap samplings. 
The dashed line is the one-to-one relation and 
 the solid line shows the linear fit to the spectroscopically identified member fraction with $P_{mem}$. 
The dotted line displays a similar relation derived from the HeCS-red sample \citep{Rines17}. }
\label{pfrac}
\end{figure}

Figure \ref{pfrac} shows 
 the spectroscopically identified member fractions as a function of $P_{mem}$. 
A large fraction, but not all of the objects with $P_{mem} > 0.9$ 
 tend to be spectroscopically identified members. 
We examine the spectroscopically identified member fractions at 
 various clustercentric radii within different magnitude ranges, 
 but the overall fractions have little dependence on these observables. 
The spectroscopically identified member fractions are also insensitive to the 
 redshift and the richness of clusters. 

We fit the spectroscopically identified member fraction with a simple linear relation, 
\begin{equation}
f_{spec-mem} = (0.25 \pm 0.02) + (0.66 \pm 0.02) P_{mem}. 
\label{eq_pfrac}
\end{equation} 
This relation provides a spectroscopically determined correction 
 to the redMaPPer membership probability. 
It is interesting that at the lowest redMaPPer probabilities, 
 there are more real spectroscopic members ($\sim 14$\%) than the redMaPPer algorithm would suggest.   

We compare the $f_{spec-mem}$ from the HectoMAP-red sample 
 to a similar relation derived from the HeCS-red sample \citep{Rines17}. 
The HeCS-red sample includes 23 high-richness ($\lambda \geq 64$) redMaPPer clusters 
 in the redshift range $0.08 < z < 0.25$. 
\citet{Rines17} examine the spectroscopically identified member fraction of the HeCS-red sample
 based on extensive SDSS and MMT/Hectospec redshift data. 
The blue dotted line in Figure \ref{pfrac} shows 
 the linear relation for the HeCS-red sample; this relation is much shallower than for the HectoMAP-red sample. 
In the HeCS-red sample, 
 fewer high $P_{mem}$ galaxies are spectroscopically identified members and 
  more low $P_{mem}$ galaxies are spectroscopically identified members. 

The HeCS-red clusters include 
 the highest richness systems at low redshift; in contrast,  
 the HectoMAP-red clusters are low richness systems covering a much wider redshift range. 
In its redshift range $0.08 < z < 0.25$, 
 the redshift survey for the HeCS-red has more complete ($\sim 90\%$) coverage of the red sequence,
 but reaches only $r \leq 20.5$.  
The $r-i$ cut in HectoMAP leads to undersampling of the red sequence 
 in this redshift range.   
In addition, 
 \citet{Rines17} use the caustic technique \citep{Diaferio97, Diaferio99, Serra13} 
 for identifying spectroscopic members. 
This approach is more stringent than our coarse membership determination. 
These substantial differences in samples and member identification probably explain 
 the differences in the spectroscopically identified member fractions.  

\subsection{Richness of HectoMAP-red Clusters}\label{richness}

Spectroscopy of the HectoMAP-red clusters 
 confirms the identification of most redMaPPer systems. 
Ninety percent or more of these systems show a concentration in the R-v diagram, 
 and the R-v diagram identifies more than 10 members. 
Overall the redMaPPer catalog has impressive purity:
$\gtrsim$ 90\% of the candidate systems are condensations in redshift space.
Based on the spectroscopy, 
 we refine the cluster mean redshift and richness. 

\begin{figure}
\centering
\includegraphics[scale=0.49]{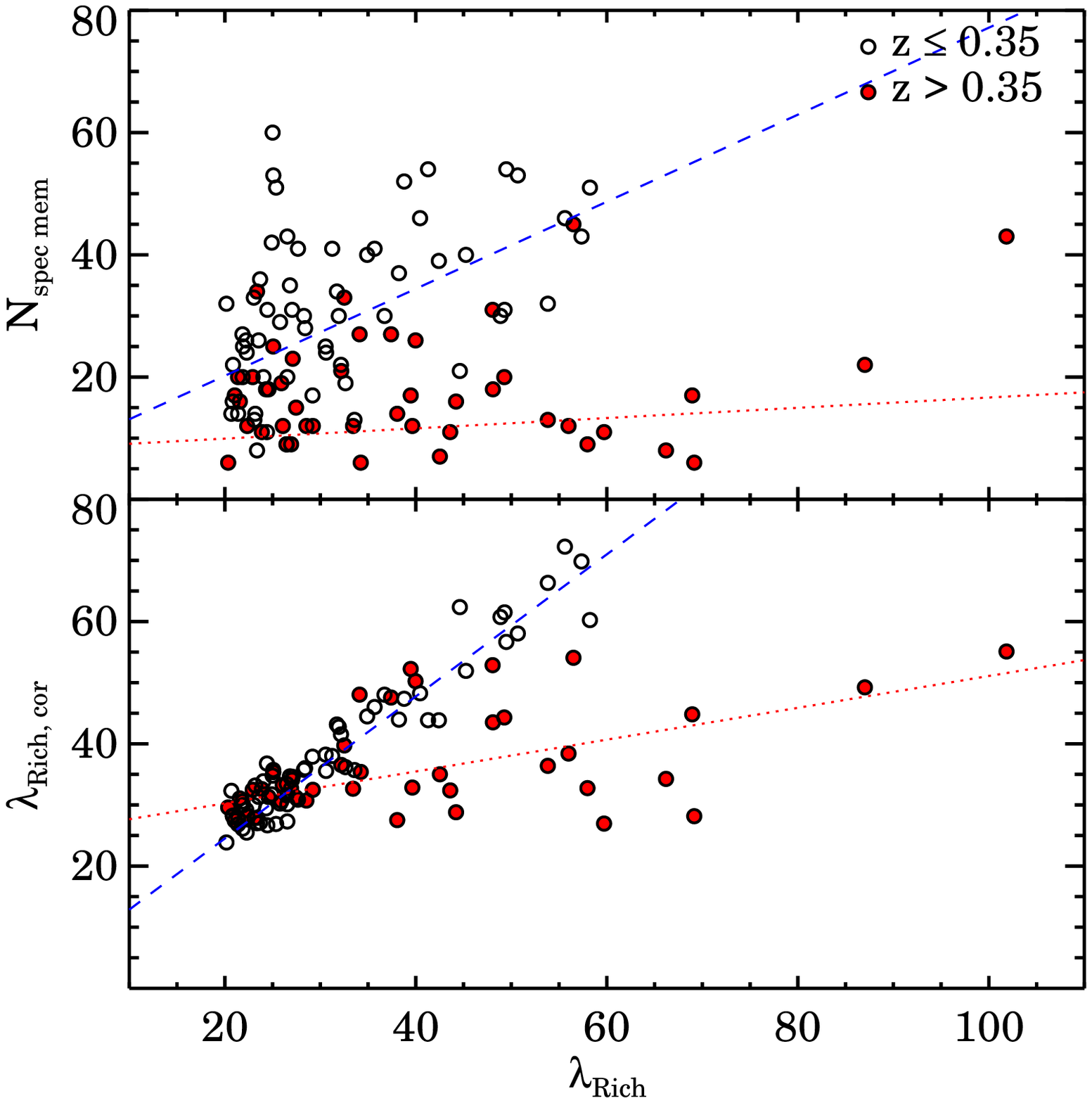}
\caption{Number of spectroscopically identified members of redMaPPer clusters as a function of richness.
The open and filled circles indicate HectoMAP-red clusters at $z \leq 0.35$ and $z > 0.35$, respectively.
The dashed and dotted lines are 
 linear fits for the lower and the higher redshift samples, respectively. }
\label{spmem}
\end{figure}

The upper panel of Figure \ref{spmem} shows 
 the number of spectroscopically identified members in the HectoMAP-red clusters 
 as a function of redMaPPer richness, $\lambda_{rich}$. 
We next test the correlation between the number of spectroscopically identified members 
 and the redMaPPer richness ($\lambda_{rich}$). 
The Pearson correlation coefficient ($0.13 \pm 0.10$) is very low.  
If we divide the samples at $z = 0.35$, 
 where redMaPPer identifies clusters based on a different color-magnitude domain, 
neither sample shows a significant correlation ($0.49 \pm 0.11$ and $0.21 \pm 0.15$, respectively).  
 
The 10 HectoMAP-red clusters with fewer than 10 spectroscopic members 
 cover a wide redMaPPer richness range $20 < \lambda_{rich} < 70$. 
Some of these systems may be poor groups. 
The HSC image of one of these candidate systems, HM13503, shows 
 that only a few ($\sim 5$) redMaPPer candidates members are luminous ellipticals.
Eight high redshift ($z > 0.35$) candidates with fewer than 10 spectroscopic members
 may be poorly sampled;
 many members have $r > 21.3$. 

Based on the spectroscopy, 
 we construct a corrected redMaPPer richness ($\lambda_{rich, cor}$).
To compute this corrected richness, 
 we use spectroscopically identified members that also have a redMaPPer membership probability ($P_{mem} > 0$).
At redshift $z > 0.3$, 
 nearly all spectroscopically identified members have a redMaPPer membership probability;
 at $z \lesssim 0.4$, the median fraction is $\sim 70\%$. 
The global fraction of spectroscopically identified members without a redMaPPer $P_{mem}$ is $\sim 25\%$. 

The corrected richness is
\begin{equation}
\lambda_{rich, cor} = \sum f_{spec-mem} = \sum (0.25 + 0.66 P_{mem}). 
\end{equation} 
In equation \ref{eq_pfrac},
 we use only redMaPPer candidate members with $r_{petro, 0} < 21.3$, 
 the limiting apparent magnitude of the HectoMAP survey. 
The corrected richness indicates 
 the total number of redMaPPer members 
 after correction for contamination by line-of-sight objects brighter than the HectoMAP limit.  
The weighting of objects without spectroscopically confirmed membership
 reflects the original redMaPPer prescription with the correction to the probabilities 
 that we derive from spectroscopy. 
We have checked that the correction to the redMaPPer membership probability (Figure \ref{pfrac}) 
 is insensitive to apparent magnitude, redshift, and color. 

The lower panel of Figure \ref{spmem} displays 
 the corrected richness as a function of the original redMaPPer richness ($\lambda_{rich}$). 
For the overall and high-z ($z > 0.35$) samples, 
 the corrected richness is not tightly correlated with the original redMaPPer richness; 
 the Pearson correlation coefficient is $0.57 \pm 0.08$. 
The incompleteness of our spectroscopic sample at $21.3 < r < 22$, 
 where most redMaPPer members in the high-z samples appear,
 may produce the lack of correlation. 
In contrast, 
 the lower redshift sample shows a significant correlation with a coefficient of $\sim 0.99$.
For low-z samples, 
 the HectoMAP redshift survey covers a significant fraction of redMaPPer members.
The linear fit between $\lambda$ and $\lambda_{rich, cor}$ for 
 the $z \leq 0.35$ HectoMAP-red clusters (after $2\sigma$ clipping) is 
\begin{equation}
\lambda_{rich, cor} =  (1.3 \pm 7.8) + (1.2 \pm 0.2) ~\lambda_{rich}. 
\end{equation}
  
\begin{figure}
\centering
\includegraphics[scale=0.49]{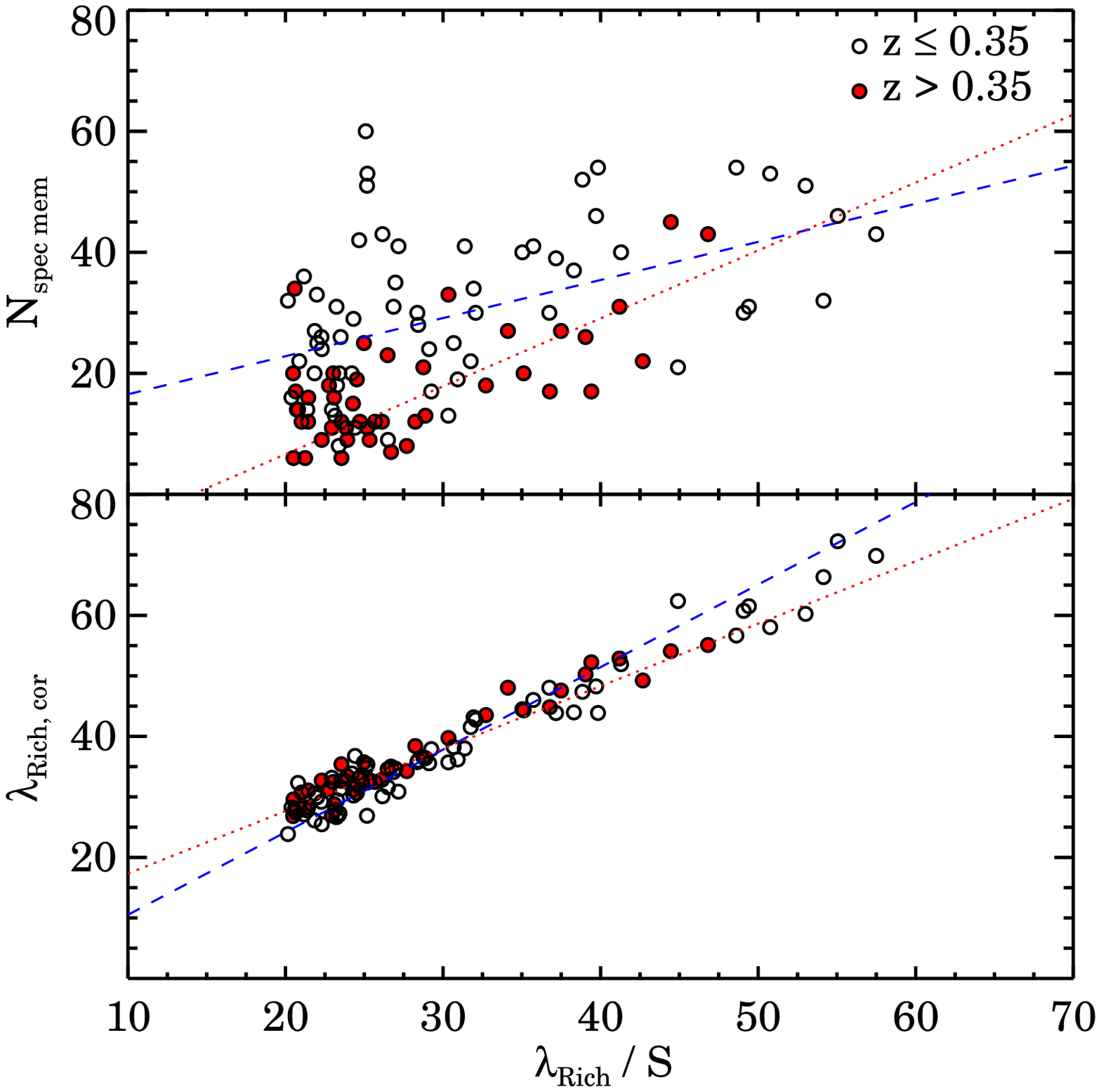}
\caption{Number of spectroscopically identified members of redMaPPer clusters 
 as a function of scaled richness ($\lambda_{rich}/S$).
The open and filled circles indicate HectoMAP-red clusters at $z \leq 0.35$ and $z > 0.35$, respectively. 
The dashed and dotted lines are 
 linear fits for the lower and the higher redshift samples, respectively. }
\label{spmem_scl}
\end{figure}
    
Figure \ref{spmem_scl} is similar to Figure \ref{spmem}, 
 but for the scaled redMaPPer richness ($\lambda_{rich}/S$).  
The scaled richness indicates 
 the effective number of redMaPPer members 
 brighter than the limiting magnitude of the survey ($r \sim 22$). 
The scaled richness corrects for the geometric survey mask 
 (including star holes and survey boundaries) within the survey area, 
 or the missing galaxies below the redMaPPer magnitude limit $i < 21.0$
 \citep{Rykoff14, Rozo15b, Rykoff16}.

The scaled richness shows a weak correlation 
 with the number of spectroscopically identified members in both the low and high redshift samples. 
The Pearson correlation test yields marginal correlation coefficients:
 $0.48 \pm 0.11$ for $z \leq 0.35$ and $0.65 \pm 0.12$ for $z > 0.35$. 
The dashed and dotted lines in Figure \ref{spmem_scl} indicate 
 linear fits for the low and high redshift samples, respectively. 
However, the spectroscopically corrected richness is tightly correlated with the scaled richness, 
 with high correlation coefficients ($0.97 \pm 0.03$ and $0.97 \pm 0.03$). 
We derive linear fits between 
 the spectroscopically corrected richness and the redMaPPer scaled richness ($\lambda_{rich}/S$) for both redshift ranges
\begin{equation}
\lambda_{rich, cor} = (-3.1 \pm 7.4) + (1.4 \pm 0.2) (\lambda_{rich}/S) \textrm{ for } z \leq 0.35, 
\end{equation}
and 
\begin{equation}
\lambda_{rich, cor} = ( 7.0 \pm 7.5) + (1.0 \pm 0.2) (\lambda_{rich}/S) \textrm{ for } z > 0.35. 
\end{equation}

Several studies suggest that 
 the redMaPPer $\lambda_{rich}$ is a mass proxy \citep{Rozo15a, Rozo15b, Simet17}.
However, HectoMAP spectroscopy suggests that 
 $\lambda_{rich}$ is not tightly correlated with the spectroscopically corrected richness
 at $z > 0.35$. 
In this redshift range, the redMaPPer scaled richness correction $S$ is significant.   
Extension of the spectroscopy to fainter magnitudes for 
 a sufficiently large sample of photometrically identified clusters would substantially improve
 the calibration of $\lambda_{rich}$ as a mass proxy, in particular at $z > 0.35$.

\subsection{Completeness of HectoMAP-red Clusters}\label{comp}

\begin{figure*}
\centering
\includegraphics[scale=0.41]{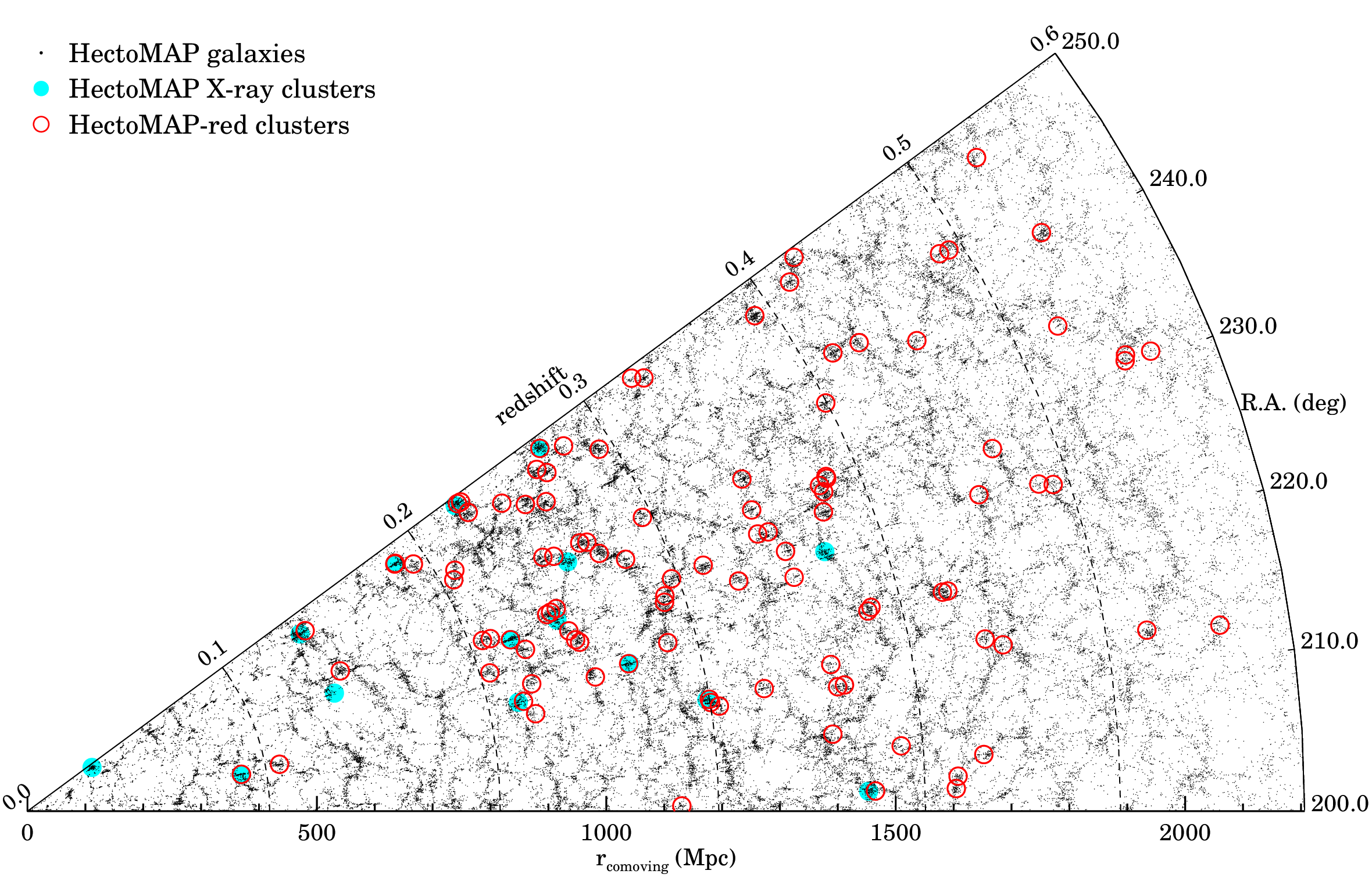}
\caption{Cone diagram for HectoMAP projected on the R.A. plane. 
Black dots show individual galaxies and cyan filled circles indicate the HectoMAP X-ray clusters.  
Red open circles display HectoMAP-red clusters based on the spectroscopic redshifts we determine. }
\label{cone}
\end{figure*}

Figure \ref{cone} shows the position of the HectoMAP-red clusters 
 in the cone diagram for the HectoMAP region. 
We mark the positions of the HectoMAP-red clusters (red circles) 
 based on the HectoMAP-red spectroscopic redshifts.
If we plotted the HectoMAP-red clusters based on the redMaPPer photometric redshift, 
 the systems would be offset from the galaxy over-densities along the line-of-sight.
 
Figure \ref{cone} also shows that 
 some of the obvious dense HectoMAP regions
 do not have HectoMAP-red clusters associated with them. 
In some of these regions, fingers in redshift space are apparent. 
For example, 
 massive clusters associated with X-ray emission (cyan circles, \citealp{Sohn17c})
 are missing from the HectoMAP-red cluster sample. 
 
We can obtain some estimate of the completeness of the redMaPPer cluster survey 
 by comparing it with the HectoMAP X-ray clusters \citep{Sohn17c} that represent some of 
 the most massive systems in the HectoMAP field with $M_{200} \gtrsim 2 \times 10^{13} M_{\odot}$. 
The X-ray clusters are identified by a co-identification method
 based on a friends-of-friends algorithm applied to HectoMAP 
 along with X-ray source detection from the ROSAT all sky survey data. 
There are 15 HectoMAP X-ray clusters
 with $0.03 < z < 0.40$. 
All of these clusters have at least 18 spectroscopic members; 
 the richest of them has 218 spectroscopically confirmed members.
   
We identify
 HectoMAP-red cluster counterparts to the X-ray systems 
 if they are within 1.5 Mpc and $|\Delta cz / (1 + z_{cl})| < 1000~\kms$. 
We set the 1.5 Mpc criterion to reflect the redMaPPer limiting size (Figure \ref{rv}).
Eleven of the HectoMAP X-ray clusters have HectoMAP-red cluster counterparts 
 in the public redMaPPer catalog. 
One HectoMAP X-ray cluster is at $z = 0.03$, below the redMaPPer survey limit. 
The three missing X-ray clusters are massive systems with large velocity dispersion 
 ($\gtrsim 450$ km s$^{-1}$, \citealp{Sohn17c}). 
The dynamical masses of two of the missing X-ray clusters are larger than 
 the effective mass cut ($\gtrsim 1.24 \times 10^{14} M_{\odot}$, Simet et al. 2017) of 
 the public redMaPPer catalog with $\lambda > 20$. 
These X-ray clusters show obvious $g-r$ or $r-i$ red sequences (Sohn et al. 2017 submitted). 
These missing X-ray clusters ($20 \pm 11\%$) are listed in a private redMaPPer catalog 
 with a lower richness threshold $\lambda > 5$ (E.Rozo, private communication). 
Tests of photometric catalogs against all sky X-ray data with deeper and 
 better resolution from e-ROSITA combined with a dense redshift survey for 
 some significant sample can provide a much more robust foundation for 
 cosmological applications than available currently.
   
\section{SUMMARY}

HectoMAP is a dense and a complete redshift survey 
 covering $\sim 53$ deg$^{2}$ and a redshift range $z < 0.6$. 
The survey is dense enough over a significant area to test
 various cluster identification techniques 
 based on photometric data. 
Surprisingly, the number of photometrically cluster candidates with redshift $z \lesssim 0.6$ in the HectoMAP region
 varies from 104 to 544 among various catalogs. 
 
We examine the 104 redMaPPer cluster candidates with $0.08 < z < 0.6$ in the HectoMAP region, 
 {\it i.e.} the HectoMAP-red clusters.
Although the redMaPPer catalog has been widely tested with multi-wavelength data, 
 the HectoMAP-red clusters are unique in testing 
 the robustness of the full richness range to the redMaPPer redshift range. 
The HectoMAP-red cluster sample complements the HeCS-red cluster sample \citep{Rines17}. 
The {\it Subaru}/Hyper Suprime-Cam (HSC) imaging archive includes 15 HectoMAP-red clusters. 
In the HSC images, most systems are apparent clusters with an obvious BCG near the cluster center.
 
The HectoMAP redshift survey yields a fairly complete ($> 60\%$) sample of redshifts for 
 redMaPPer candidate members in the HectoMAP-red clusters. 
We determine the cluster central redshift and the spectroscopic cluster membership
 based on redshifts of individual member candidates. 
The redMaPPer algorithm identifies 
 $16 - 107$ member candidates with $r \leq 21.3$ for each HectoMAP-red cluster; 
 we identify $\sim 6 - 60$ spectroscopic members. 
We include 3547 redshifts for member candidates 
 listed in the redMaPPer catalog. 
  
The redMaPPer central galaxies are identical to 
 the spectroscopically determined brightest cluster galaxies (BCGs) 
 in 76 of the 104 clusters. 
The HectoMAP redshift survey does not include redshifts of central galaxies 
 for four systems. 
In 24 systems ($\sim 23\%$), 
 the central galaxies are not spectroscopic BCGs;
 sometimes they are not even cluster members. 
This fraction of offset BCG is consistent with 
 the redMaPPer estimate of the central galaxy misidentifications ($\sim 15 - 20\%$). 

We estimate the spectroscopically identified member fraction 
 among redMaPPer member candidates. 
Overall, $\sim60\%$ of redMaPPer member candidates are spectroscopically identified members. 
Interestingly, 
 not all HectoMAP-red member candidates with the highest redMaPPer membership probability ($P_{mem} > 0.9$) 
 are spectroscopically identified members. 
In fact,  
 $\sim 15\%$ of the lowest $P_{mem}$ galaxies ($P_{mem} < 0.1$) are spectroscopically identified members. 
Based on the spectroscopically identified member fraction,
 we provide a correction function for the redMaPPer membership probability. 
   
We compare the photometrically estimated redMaPPer richness, $\lambda_{rich}$, 
 with the spectroscopic richness. 
The richness, $\lambda_{rich}$, is not well-correlated with the spectroscopic richness at redshift $z > 0.35$. 
However, the spectroscopic richness correlates well with the scaled richness ($\lambda_{rich}/S$) 
 in the redMaPPer catalog throughout the HectoMAP redshift range. 
 
The HectoMAP redshift survey demonstrates 
 that $\sim 10\%$ of the HectoMAP-red clusters are possibly loose groups
 with fewer than 10 spectroscopically identified members. 
More importantly, 
 the redMaPPer algorithm fails identify all of the massive clusters in the HectoMAP region.
For example, 
 $\sim 20 \pm 11\%$ (3 systems) of the well-populated massive clusters associated with ROSAT X-ray emission 
 are not recovered by redMaPPer. 
Further tests of photometric cluster catalogs 
 against a dense redshift and deeper X-ray data 
 are crucial for providing a robust list of clusters 
 for studying formation and evolution of large-scale structures and for limiting the cosmological parameters. 
 
\acknowledgments

The authors are grateful to the referee, 
 Eduardo Rozo, for his comments that improved the clarity of the paper.
JS gratefully acknowledges the support of a CfA Fellowship. 
The Smithsonian Institution supports MJG. 
AD acknowledges support from the INFN grant InDark. 
We thank David Reiman for assistance during the early stages of this work.
We acknowledge Susan Tokarz for reducing the spectroscopic data and 
 Perry Berlind and Mike Calkins for assisting with the observations. 
We also thank the telescope operators at the MMT and Nelson Caldwell
 for scheduling Hectospec queue observations.
We thank the HSC help desk team, especially Michitaro Koike and Sogo Mineo, for making the useful tools available.
This research has made use of NASAs Astrophysics Data System Bibliographic Services. 
 
The Hyper Suprime-Cam (HSC) collaboration includes the astronomical communities of Japan and Taiwan, and Princeton University. 
The HSC instrumentation and software were developed by the National Astronomical Observatory of Japan (NAOJ), 
 the Kavli Institute for the Physics and Mathematics of the Universe (Kavli IPMU), 
 the University of Tokyo, the High Energy Accelerator Research Organization (KEK), 
 the Academia Sinica Institute for Astronomy and Astrophysics in Taiwan (ASIAA), and Princeton University. 
Funding was contributed by the FIRST program from Japanese Cabinet Office, 
 the Ministry of Education, Culture, Sports, Science and Technology (MEXT), 
 the Japan Society for the Promotion of Science (JSPS), Japan Science and Technology Agency (JST), 
 the Toray Science Foundation, NAOJ, Kavli IPMU, KEK, ASIAA, and Princeton University. 

This paper makes use of software developed for the Large Synoptic Survey Telescope. 
We thank the LSST Project for making their code available as free software at  http://dm.lsst.org

The Pan-STARRS1 Surveys (PS1) have been made possible through contributions of the Institute for Astronomy, 
 the University of Hawaii, the Pan-STARRS Project Office, the Max-Planck Society and its participating institutes, 
 the Max Planck Institute for Astronomy, Heidelberg and the Max Planck Institute for Extraterrestrial Physics, Garching,
 the Johns Hopkins University, Durham University, the University of Edinburgh, Queen’s University Belfast, 
 the Harvard-Smithsonian Center for Astrophysics, the Las Cumbres Observatory Global Telescope Network Incorporated, 
 the National Central University of Taiwan, the Space Telescope Science Institute, 
 the National Aeronautics and Space Administration under Grant No. NNX08AR22G issued through 
 the Planetary Science Division of the NASA Science Mission Directorate, 
 the National Science Foundation under Grant No. AST-1238877, the University of Maryland, 
 and Eotvos Lorand University (ELTE) and the Los Alamos National Laboratory. 
 
\facility{{\it Facility } : MMT (Hectospec)}


\end{document}